\documentclass{cambridge6A}
 \usepackage{graphicx}
 \usepackage{amsmath}
 \usepackage{amsfonts}

\begin{document}

\chapterauthor{Veronika E Hubeny, Shiraz Minwalla, Mukund Rangamani}

\chapter{The fluid/gravity correspondence}

\copyrightline{Chapter of the book \textit{Black Holes in Higher Dimensions} to
be published by Cambridge University Press (editor: G. Horowitz)}

\contributor{Veronika E Hubeny \affiliation{Durham Univeristy}}
\contributor{Shiraz Minwalla \affiliation{Tata Institute of Fundamental Research}}
\contributor{Mukund Rangamani\affiliation{Durham Univeristy}}

%
 
\section{Introduction}
\label{s:intro}

In this chapter we will study a particular long wavelength limit of 
Einstein's equations with a negative cosmological constant in $d+1$ dimensions.
 In such a limit we find that Einstein's equations reduce to the equations of fluid dynamics 
(relativistic generalizations of the famous Navier-Stokes equations) in 
$d$ dimensions. While the motivation for  our study lies within the AdS/CFT correspondence of string theory, the fluid/gravity correspondence stands on its own and can be viewed as a map between two classic dynamical systems.  

\subsection{Prelude: CFT stress tensor dynamics from gravity }
\label{s:prelude}

An important consequence of the AdS/CFT correspondence [Ch.AdSCFT] is that the dynamics of the stress(-energy-momentum) tensor in a large class of $d$-dimensional strongly coupled quantum field theories is governed by the dynamics of Einstein's equations with negative cosmological constant in $d+1$ dimensions. To begin with, we shall try to provide the reader with some intuition for this statement and argue that searching for a tractable corner of this connection leads one naturally to the fluid/gravity correspondence.

In its most familiar example, the AdS/CFT correspondence [Ch.AdSCFT]  
asserts that $SU(N)$  ${\cal N}=4$ Super Yang-Mills (SYM) theory is dual to Type IIB string theory  on AdS$_5 \times {\bf S}^5$.  It has long been known that in the 't Hooft limit, 
which involves taking $N\to \infty$ keeping the coupling $\lambda$ fixed, the gauge theory becomes effectively classical. However, it was widely believed that  for any non-trivial gauge theory  the resulting classical system would be too complicated to be tractable. The remarkable observation of Maldacena in 1997 was that this field theory intuition is spectacularly wrong. Indeed, not only is the  classical system governing ${\cal N}=4$ SYM tractable, it is actually a well known theory, viz., classical Type IIB string theory.

Now, even classically, string theory has complicated dynamics; however in the strong gauge coupling ($\lambda \to \infty$) regime, it reduces to the dynamics of Type IIB supergravity (by decoupling the massive string states). More interestingly, Type IIB supergravity on AdS$_5\times {\bf S}^5$ admits several consistent truncations. The simplest and most universal of these is the truncation to Einstein's equations with negative 
cosmological constant,
\begin{equation}\label{Eeq}
 E_{\mu\nu} \equiv R_{\mu\nu} - \frac{1}{2} \, R \, g_{\mu\nu} + \Lambda  \, g_{\mu\nu} =0 \ , \qquad \Lambda \equiv -\frac{d \, (d-1)}{2\, R_\text{AdS}^2} \ .
 \end{equation}
(Note that the  AdS curvature radius  $R_\text{AdS}$ can be scaled away by a change of units ; we therefore set $R_\text{AdS}$ to unity in the rest of this chapter). Having thus motivated the study of the most beautiful equation of physics, namely Einstein's equations of general relativity, we now confront the question: What does this imply for the field theory?

Recall that according to the AdS/CFT dictionary there is a one-to-one map between single particle states in the classical Hilbert space of string theory and  single-trace operators in the gauge theory. For instance the bulk graviton maps to the stress tensor of the boundary theory. Taking the collection of such single trace operators as a whole, one can try to formulate dynamical equations for their quantum expectation values in the field theory. While this can be done in principle,  the resulting system is non-local in terms of the intrinsic field theory variables themselves. 

However, because we can associate the quantum operators (and their expectation values) of the gauge theory  at strong coupling to the classical fields of string theory/supergravity, we know that the set of classical equations we are looking for are just the local equations of Type IIB supergravity on AdS$_5 \times {\bf S}^5$. This reduction, whilst retaining lots of interesting physics, still turns out to be too complicated from the field theory perspective. For one, the space of single trace operators is still infinite dimensional (at infinite $N$), and relatedly attempting to classify the solution space of Type IIB supergravity is a challenging problem. However, the fact that on the string side we can reduce the system to \eqref{Eeq}, implies that there is a decoupled sector of stress tensor dynamics in ${\cal N}=4$ SYM at large $\lambda$.\footnote{While this is always true in two dimensional field theories, such a decoupling is not generic in higher dimensional field theories (in fact it is not true of ${\cal N}=4$ Yang Mills at weak coupling), and is in itself a surprising and interesting 
fact about the ${\cal N}=4$ dynamics at strong coupling. } 

Actually, there is an infinite number of conformal gauge theories which have a gravitational dual that truncates consistently at the two-derivative level to Einstein's equations with a negative cosmological constant; ${\cal N}=4$ SYM theory is just a particularly simple member of this class. Thus \eqref{Eeq} describes the  {\it universal decoupled} dynamics of the stress tensor for an infinite  number of different gauge theories. In the first part of this chapter 
we will focus on the study of this universal sector. Later we will generalize to the
study of bulk equations with more fields, thereby obtaining richer dynamics at the expense of universality. 

Given this association between the dynamics of quantum field theory stress tensors to the dynamics of gravity in negatively curved backgrounds, it is natural to ask -- can we do more? Can we for instance classify all possible behaviors of stress tensors? On the gravity side we would have to classify all possible solutions to \eqref{Eeq}; this is a laudable goal and various chapters in this book are aimed at addressing this question using different approaches. We are going to focus on one that naturally follows from the basic organizing principle of physics: separation of scales. 

It is well known that in many situations in physics (as well as chemistry, biology, etc.), 
complicated UV dynamics results in relatively simple IR dynamics. Perhaps the first systematic exposition of this ubiquitous fact was in the context of finite  temperature physics. It has been known for almost 200 years now that  the dynamics of nearly equilibrated systems at high enough  temperature may be described by an effective theory called hydrodynamics.
The key dynamical equation of hydrodynamics is the conservation of the stress tensor 
\begin{equation}
\nabla_{\! a} \, T^{ab} = 0 \ ,
\label{Tcons}
\end{equation}
where $\nabla_{\! a}$ is the covariant derivative compatible with the background metric $\gamma_{ab}$ on which this fluid lives. As this equation is an autonomous dynamical system involving just the stress tensor, it should lie within the sector of universal decoupled stress tensor dynamics. 

Given that the AdS/CFT correspondence asserts that this universal sector 
is governed by \eqref{Eeq},  we are led to conclude that 
\eqref{Eeq} must, in an appropriate high temperature
and long distance limit which we refer to as the {\it long wavelength regime}, reduce to the equations of $d$-dimensional hydrodynamics. Indeed, this expectation has been independently verified  in 
\cite{Bhattacharyya:2008jc} and the resulting map between gravity and 
fluid dynamics has come to be known as the {\it fluid/gravity correspondence}.  In particular, the specific fluid 
dynamical equations, dual to long wavelength gravity in the universal sector, have been determined up to the second order in a gradient expansion (cf.\ \S\ref{s:sten2}). Given any solution to the 
these fluid dynamical equations, the fluid/gravity map  {\it explicitly} determines 
a solution to Einstein's equations \eqref{Eeq} to the appropriate 
order in the derivative expansion. The solutions in gravity are simply inhomogeneous, time-dependent black holes, with slowly varying but otherwise generic horizon profiles. 
 
The main focus of the present chapter is to explain and present the fluid/gravity map at the full non-linear level following \cite{Bhattacharyya:2008jc} and subsequent work. 
The connection between these two systems was established and extensively studied 
much earlier at the linearized level in the AdS/CFT context (following the seminal work 
\cite{Policastro:2001yc}). The first hints of the connection between fluid dynamics and gravity at the non-linear level were obtained in attempts  to construct non-linear solutions dual to a particular boost invariant flow \cite{Janik:2005zt}, which provided inspiration for the fluid/gravity map. Such a map was also suggested by the observation that the properties of large 
rotating black holes in global AdS space are reproduced by the equations 
of non-linear fluid dynamics \cite{Bhattacharyya:2007vs}. We refer the reader to \cite{Rangamani:2009xk} for a list of developments and references.

\subsection{Preview of the fluid/gravity correspondence}
\label{s:ipreview}

Having provided the reader with a broad, albeit abstract, rationale to associate the dynamics of Einstein's equations to that of a quantum field theoretic stress tensor, we now provide some specifics that set the stage for our discussion.

According to the gauge/gravity dictionary, distinct asymptotically AdS bulk geometries correspond to distinct states in the boundary gauge theory.
The  pure AdS geometry, i.e., the maximally symmetric negatively curved spacetime, corresponds to the vacuum state of the gauge theory. A large\footnote{
Recall that AdS is a space of constant negative curvature, which introduces a length scale, called the AdS scale $R_{\rm AdS}$, corresponding to the radius of curvature.  
The black hole size is then measured in terms of this AdS scale; large black holes have horizon radius $r_+  > R_{\rm AdS}$.  We will be focus on the large black hole limit $r_+  \gg R_{\rm AdS}$, and therefore consider the  {\it planar} Schwarzschild-AdS black holes.
} Schwarzschild-AdS black hole corresponds to a thermal density matrix in the gauge theory.  This can be easily conceptualized in terms of the late-time configuration a generic state evolves to: in the bulk, the combined effect of gravity and negative curvature tends to make a generic configuration collapse to form a black hole which settles down to the Schwarzschild-AdS geometry, while in the field theory, a generic excitation will eventually thermalize. 
Note that although the underlying theory is supersymmetric, the correspondence applies robustly to non-supersymmetric states such as the black holes mentioned above.  In this sense, supersymmetry is {\it not} needed for the correspondence.

On the boundary, the essential physical properties of the gauge theory state (such as local energy density, pressure, temperature, entropy current, etc.) are captured by the expectation value of the {\it boundary stress tensor}, which in the bulk is related to normalizable metric perturbations about a given state. It can be extracted via a well-defined Brown-York type procedure \cite{Balasubramanian:1999re} as we review later (see \eqref{BYstress}). 

At the risk of being repetitive we urge the reader to note the distinction between the two separate stress tensors that will enter our analysis.  In our framework, the {\it bulk}\ stress tensor appearing on the r.h.s.\ of the bulk  Einstein's equation is zero if we are only interested in the universal sub-sector discussed above.  On the other hand, the {\it boundary} stress tensor $T^{ab}$ is non-zero; it is a measure of the  normalizable fall off of the bulk metric at the boundary. Note that the boundary stress tensor does not 
curve the boundary spacetime \`a la Einstein's equations since the boundary metric $\gamma_{ab}$  is non-dynamical and fixed. We will discuss generalizations that allow for non-trivial bulk matter in \S\ref{s:extensions} when we move outside the universal stress tensor sector.

To describe gravity duals of fluid flows, a useful starting point is  the map between the boundary and bulk dynamics in global thermal equilibrium. In the field theory, one characterizes thermal equilibrium by a choice of static frame and a temperature field.  On the gravity side, the natural candidates to characterize the equilibrium solution are static (or more generally stationary) black hole spacetimes, as can be seen by demanding regular solutions with periodic Euclidean time circle. The temperature of the fluid is given by the Hawking temperature of the black hole, while the fluid dynamical velocity is captured by the horizon boost velocity of the black hole. For planar Schwarzschild-AdS black holes the temperature grows linearly with horizon size; the AdS asymptotics thus ensures  thermodynamic stability as well as providing a natural long wavelength regime.

Now let us try to gently move away from the equilibrium configuration.
Starting with the stationary black hole (namely the boosted planar Schwarzschild-AdS$_{d+1}$) solution, we wish to use it to build solutions where the fluid dynamical temperature and velocity are slowly-varying functions of the boundary directions. Intuitively, this mimics patching together pieces of black holes with slightly different temperatures and boosts in a smooth way so as to get a regular solution of \eqref{Eeq}. In order to obtain a true solution of Einstein's equations, the patching up procedure cannot be done arbitrarily; one is required at the leading order to constrain the velocity and temperature fields to obey the equations of ideal  fluid dynamics.\footnote{These constraints are actually the radial momentum constraints for gravity in AdS and imply \eqref{Tcons}. In contrast to the conventional ADM decomposition, we imagine foliating the spacetime with timelike leaves and `evolve' into the AdS bulk radially.} Further, the solution itself is corrected order 
by order in a derivative expansion, a process that likewise corrects the fluid equations. All these steps may be implemented\footnote{In the technical implementation of this program, it is important that one respect boundary conditions. We require that the 
bulk metric asymptote to $\gamma_{ab}$ (up to a conformal factor) and further be manifestly regular in the part of the spacetime outside of any  event horizon.}  in detail in a systematic boundary gradient expansion. The final output is 
a map between solutions to negative cosmological constant gravity and 
the equations of fluid dynamics in one lower dimension, i.e.  
the fluid/gravity map. 

\begin{figure}[h!]
 \begin{center}
 \includegraphics[scale=0.55]{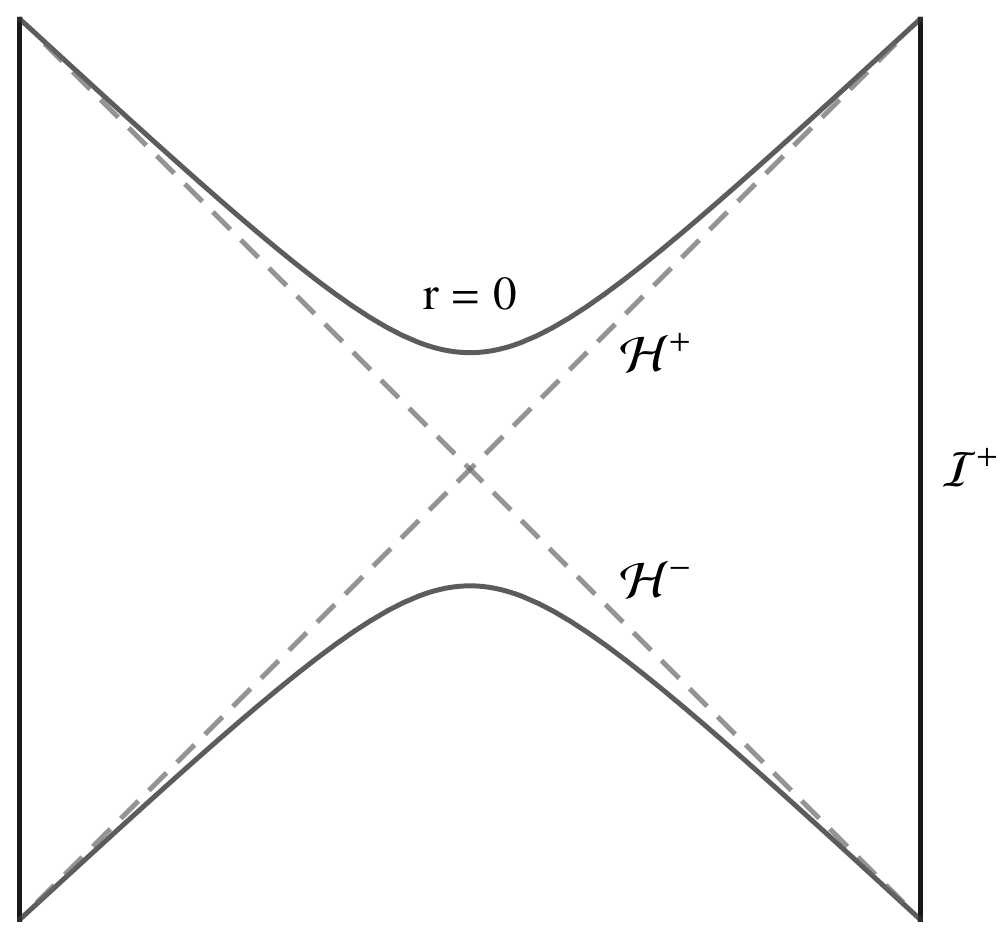}
\hspace{1.5cm}
 \includegraphics[scale=0.29]{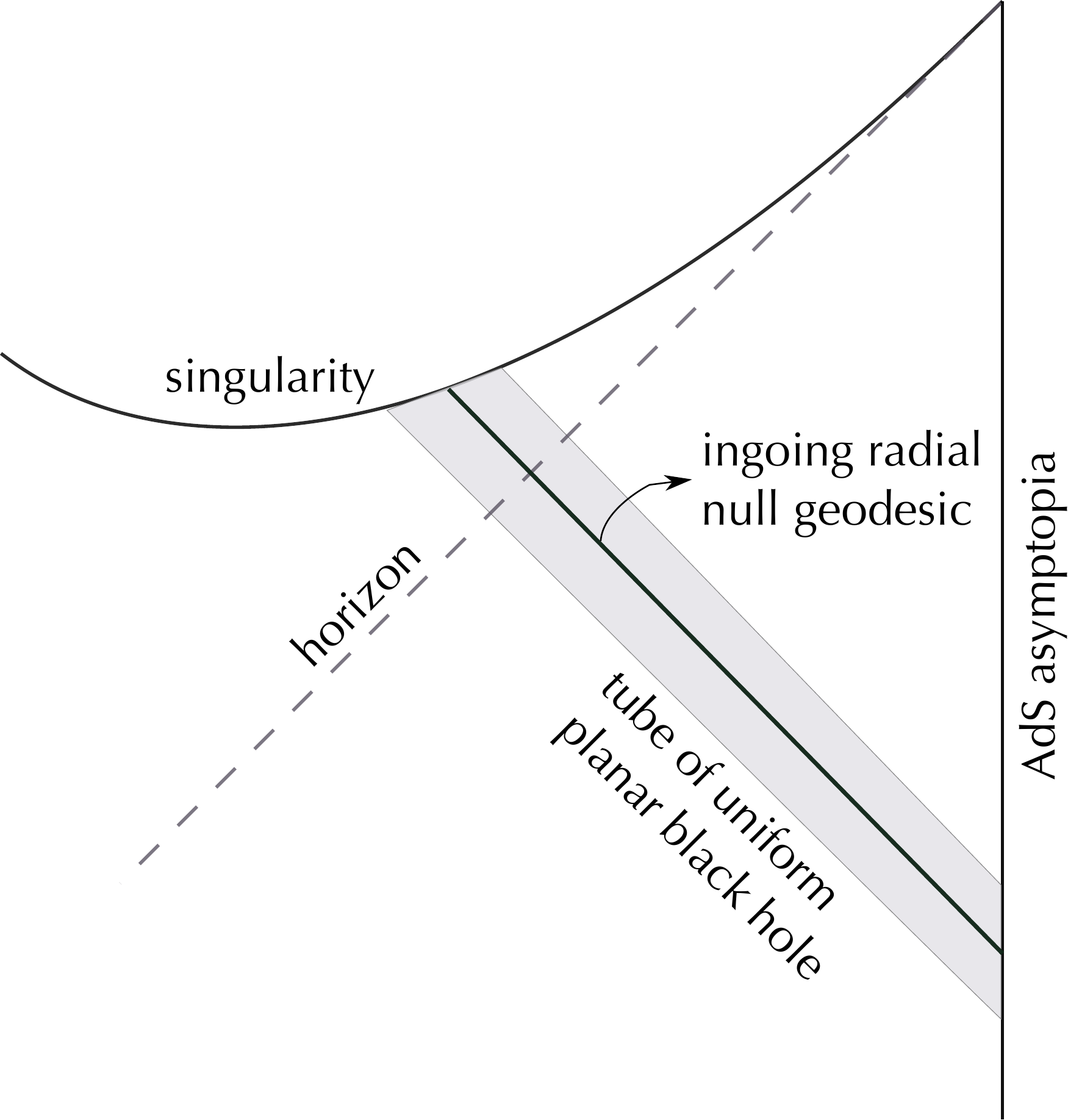} 
\end{center}
\caption{Penrose diagram of the uniform planar black hole (\ref{sads1})  and the causal structure of the spacetimes dual to fluid dynamics illustrating the tube structure. Dashed line in the second figure denotes the future event horizon, while the shaded tube indicates the region of spacetime over which the solution is well approximated by a corresponding uniform black hole.}
\label{PDtube}
\end{figure}

A noteworthy aspect of this construction is that Einstein's equations become tractable due to the long wavelength regime without losing non-linearity.  From the boundary standpoint, one encounters  domains of nearly constant fluid variables; these domains can then be extended radially from the boundary into the bulk and in each such bulk `tube', illustrated in Figure \ref{PDtube}, we are guaranteed to have a solution which is close to the equilibrium form.  Lest the reader be led astray, we should note that the solutions we construct are perturbative and hence approximate.
Nevertheless, they are `generic' slowly-varying asymptotically-AdS black hole geometries, with no Killing fields.

A remarkable outcome of the association between generic black holes and fluid flows is that it automatically provides a sensible entropy current with non-negative  divergence for hydrodynamics. On the gravitational side, entropy is naturally associated to the area of the event horizon; by pulling back this area form to the boundary, we can equip our fluid with a canonical entropy current.

We now turn to the technical aspects of the fluid/gravity map. Following a review of fluid dynamics and the perturbative construction of gravitational solutions, we finally present the main results (in particular the bulk metric and the boundary stress tensor, to second order in boundary derivatives) in \S\ref{s:secord}. The subsequent sections are devoted to describing some   implications and extensions of the basic construction.

\section{Relativistic fluid dynamics}
\label{s:fd}

To set the stage, let us start by reviewing fluid dynamics, explicating the use of gradient expansion as an organizational principle.
At high temperatures every non-trivial quantum field theory (and every 
experimentally realizable system) equilibrates into a fluid phase, i.e., 
a translationally invariant phase in which adiabatic displacement of 
neighboring elements requires no force. 

Weakly interacting fluids are composed of 
a collection of a large number of 
long lived partonic excitations which continually collide with each other. 
The time and space intervals between successive collisions of a given 
parton are called the mean free time, $t_{\rm m}$, and the mean free length, $\ell_{\rm m}$, 
respectively.\footnote{
In a relativistic system of massless particles, like 
${\cal N}=4$ SYM, $t_{\rm m} \sim \ell_{\rm m}$. We will assume this is the case
in our discussions below.} Such fluids are 
characterized by a parton density function in phase space, and the 
time evolution of this function is governed by the well known 
Boltzmann transport equations of statistical physics. These equations have 
an interesting property: arbitrary initial density functions
relax  to local thermal 
equilibrium over a time scale of order the mean free time. In other words, for $t \gg t_{\rm m}$, the parton distribution in 
momentum space approximately reduces, at every point $x$, to an
equilibrium distribution. However the parameters\footnote{
For simplicity,  in this discussion we assume that the system has 
no conserved charges other than the stress-energy-momentum tensor, and no other 
Goldstone-like light degrees of freedom. We discuss generalizations  
below.}
 characterizing this equilibrium configuration, the  
temperature field $T(x)$ and fluid velocity field $u^a(x)$, vary  on a length scale that is large compared to 
$\ell_{\rm m}$. $T(x)$ and $u^a(x)$ are the effective dynamical variables of 
the system at later times; their evolution as a function of time is governed 
by the equations of fluid dynamics. 

Now it turns out that the equations of fluid dynamics 
may also be derived in a much simpler and more general 
manner, and so apply even at strong coupling. 
The main assumption that underlies fluid  
dynamics is that systems always equilibrate locally over a finite 
time scale that we continue to refer to as $t_{\rm m}$. While this assumption is 
true of the Boltzmann transport equations, it is believed to hold more 
generally also for strongly coupled fluids. It follows immediately 
that $T(x)$ and $u^a(x)$ are the effective variables for dynamics at length 
and time scales large compared to $\ell_{\rm m}$ and $t_{\rm m}$. As we will now see, 
the equations of fluid dynamics follow inevitably out of this conclusion. 

\subsection{The equations of fluid dynamics and constitutive relations}

The stress tensor in any $d$-dimensional quantum field theory on a background with metric $\gamma_{ab}$ obeys  the $d$ conservation equations 
\begin{equation}
\nabla_{\! a} T^{ab}=0.
\label{stc}
\end{equation}
These equations do not constitute a well defined initial value problem 
for the stress tensor in general as, in $d \geq 2$, we have more variables 
(the $\frac{1}{2}\, d(d+1)$ independent components of the 
stress tensor) than equations.\footnote{
For the special  case of conformal field theories the number of variables is reduced by  one, as the trace of the stress tensor vanishes. In this case, at $d=2$,  we have as many variables as equations. This observation underlies the  special simplicity of CFTs in $d=2$.}
In the fluid dynamical limit, however, the stress tensor is determined as a 
function of $d$ variables, $T(x)$ and $u^a(x)$. Consequently, 
\eqref{stc}, supplemented with a formula for $T^{ab}$ as a function 
of thermodynamical fields, constitute a complete set of dynamical 
equations. These are the equations of fluid dynamics. 

A constitutive relation that expresses $T^{ab}$ as a function of $T(x)$, $u^a(x)$, 
and their derivatives turns \eqref{stc} into a concrete set of fluid dynamical 
equations. In thermal equilibrium the stress tensor $T^{ab}$ 
is given by 
\begin{equation} \label{Teq}
T^{ab}= \left( P + \rho    \right)\,  u^a \, u^b + P\,  \gamma^{ab}
\end{equation}
where $P$ is the pressure of the fluid and $\rho$ is its energy density. Recall 
that both $\rho$ and $P$ are known functions of temperature 
(determined by the thermodynamic equation of state of the fluid). 
For a fluid in {\it local} thermal equilibrium, 
\eqref{Teq} generalizes to 
\begin{equation} \label{const}
T^{ab}(x)= \left[ P(x) + \rho(x) \right] u^a(x) \, u^b(x) 
+ P(x) \, \gamma^{ab} + \Pi^{ab}(x)
\end{equation}
where $P(x)=P(T(x))$, $\rho(x)=\rho(T(x))$ and 
$\Pi^{ab}(x)$ represents the contributions of derivatives of $T(x)$ 
and $u^a(x)$ to the stress tensor. This dissipative part $\Pi^{ab}$ may then be 
expanded as 
\begin{equation}\label{pexp}
\Pi^{ab}= \sum_{n=1}^\infty \ell_{\rm m}^n \, \Pi_{(n)}^{ab}
\end{equation}
where $\Pi_{(n)}^{ab}$ is defined to be of $n^{\rm th}$ order in derivatives of the 
fluid dynamical fields. Note that magnitude of $\Pi_{(n)}^{ab}$ 
relative to the ideal fluid stress tensor is approximately $(\ell_{\rm m}/L)^n$ 
where $L$  characterizes the length scale of
variation of the temperature and velocity fields; consequently terms at 
higher values of $n$ are increasingly subdominant in the fluid dynamical limit.

The explicit form of the functions $\Pi_{(n)}^{ab}$ can only be 
derived from a detailed study of the dynamics of the specific system. However
the allowed forms for constitutive relations are significantly constrained 
by symmetry and other general considerations. At first order, for instance,
it is possible to assert on very general grounds that
\begin{equation}\label{constraints}
\begin{split}
 \Pi_{(1)}^{\langle a b \rangle} \equiv
P^a_{\ c}\,  P^b_{\ d}\, \Pi_{(1)}^{c d}  - \frac{1}{d-1}\,P^{ab}\, P_{cd}\, \Pi_{(1)}^{cd} &
= -2\,\eta\, \sigma^{ab}  \\
\frac{1}{d-1}\, \Pi_{(1)}^{ab} P_{ab} - \frac{\partial P}{\partial\rho} \, ( u_a \,u_b \,\Pi_{(1)}^{ab})& = - \zeta \, \theta 
\end{split}
\end{equation}
where 
\begin{equation}\label{pdef}
	P^{ab} \equiv u^{a}u^{b} + \gamma^{ab}
\end{equation} 
is the projector onto space in the local fluid rest frame,
\begin{equation}
\label{E:sigmadef}
	\sigma^{ab} =
	\nabla^{\langle a } \,  u^{ b \rangle}
	\equiv
	P^{a c}\, P^{b d}\,  
	\left( \nabla_{\! (c} \,  u_{d)} - \frac{1}{d-1}\, 
	P_{c d} \,\theta\right)
\end{equation}
is the fluid shear tensor,  $\theta \equiv \nabla_{\! c} \, u^c $ is the expansion, and the brackets around the indices ${\langle a b \rangle}$ denote the symmetric transverse traceless part of the expression. Here  $\eta$ and $\zeta$ are arbitrary functions of the temperature, referred to as the shear and bulk viscosity, respectively. 

The  equation \eqref{constraints} is a physically complete specification of the 
constitutive relations at first order, even though it leaves $P_{\ c}^a \, \Pi_{(1)}^{c d} \, u_d$ and  one linear combination of 
$\Pi_{(1)}^{ab}\, P_{ab}$ and $u_a\, u_b\, \Pi_{(1)}^{ab}$ unspecified.
This is because $T(x)$ and $u^a(x)$ have no intrinsic 
definition out of equilibrium. All equations of fluid dynamics must be `field 
redefinition invariant' (invariant under redefinitions of $T(x)$ and $u^a(x)$ 
that reduce to identity in equilibrium), and it turns out that the l.h.s.\ of 
\eqref{constraints} are the only field redefinition invariant data 
in $\Pi_{(1)}^{ab}$. The other components of $\Pi_{(1)}^{ab}$ can be modified at 
will by an appropriate field redefinition, and have no physical significance. 

The r.h.s.\ of the two equations in \eqref{constraints} represent the most general 
inequivalent `tensor' and `scalar' data that can be constructed 
from a single derivative of fluid dynamical fields compatible with the conservation equation at first order which imply that $u^a \, \nabla_{\! a} \, T\propto \theta$.

It is sometimes convenient to fix the field redefinition ambiguity by 
giving the fields $u^a(x)$ and $T(x)$ unambiguous (but arbitrary) meaning. 
In the so-called `Landau Frame' this is achieved by asserting that, at each 
point,
\begin{equation}\label{landauframe}
T^{a}_{\ b}(x) \, u^b(x) = -\rho(x) \, u^a(x) \ .
\end{equation}
This relation defines $u^a$ by identifying it with the unique timelike 
eigenvector of the stress tensor at any point, and defines the temperature 
by identifying the corresponding eigenvalue with the energy density.%
\footnote{
Note that the equations \eqref{landauframe} are true in 
equilibrium.}
In the Landau frame, which we adopt for most of this chapter, \eqref{constraints} simplify to 
\begin{equation}\label{constn}
\Pi_{(1)}^{ab}= -2\, \eta \,\sigma^{ab} -\zeta \, \theta \,  P^{a b} \ .
\end{equation}

As the equations of fluid dynamics are both local and 
thermodynamical in nature, they must respect a local form of the second law of thermodynamics. It follows that the equations
of fluid dynamics must be accompanied by an entropy current whose divergence 
is pointwise non-negative in every conceivable fluid flow.  At first order, for a charge-free fluid, the constraints imposed by this requirement are a 
relatively mild set of inequalities on $\eta$ and $\zeta$: 
 It turns out that the entropy  current is constrained to take the form 
\begin{equation}\label{canent} 
J_s^a = s \,u^{a} - \frac{1}{T}\; u_{b} \,\Pi_{(1)}^{a b}  \,,
\end{equation}
where $s$ is the entropy density. It is possible to demonstrate that 
the current in \eqref{canent} is field redefinition invariant to first 
order.  Note that the second term on the r.h.s.\ of \eqref{canent} vanishes in the Landau frame. Using the Euler relation ($\rho+P=s \, T$) and the Gibbs-Duhem relation ($dP=s \, dT$) of thermodynamics along with the  equations of motion, it follows that the divergence of this entropy current is given by
\begin{equation}\label{diventcan}
 \nabla_{\! a} J_s^a = -\nabla_{a} \left( \frac{u_{b}}{T} \right) \Pi_{(1)}^{a b} \,.
\end{equation}
Using \eqref{constraints} (or more simply \eqref{constn} in the Landau frame), 
it is easy to verify that positivity of the entropy current requires that 
$\eta \geq 0$ and $\zeta \geq 0$. At higher orders in the derivative expansion
(for uncharged fluids) and even at first order for more complicated fluids 
(e.g.\ charged fluids and superfluids) the requirement of positivity of the 
entropy current imposes more than a set of inequalities on transport 
coefficients; it forces linear combinations of otherwise 
arbitrary transport coefficients to vanish \cite{Bhattacharya:2011tr}. 

In the rest of this chapter we will be especially interested in the fluid
dynamics of conformal field theories. These theories enjoy three key 
simplifications. First, as they have no dimensionless parameters, the 
dependence of all physical quantities (e.g.\ $P$, $\rho$, $\eta$) on temperature
follows on dimensional grounds. In particular, 
\begin{equation}
P=\alpha\,  T^d, ~~~\rho=(d-1)\, \alpha  \, T^d, ~~~\eta= \eta' \, T^{d-1}
\label{cftPetc}
\end{equation}	
where $\alpha$ and  $\eta'$ are dimensionless constants. 
Second,  the stress tensor in any CFT is  necessarily traceless. It in particular follows from this condition that  $\zeta=0$. Finally, the stress tensor in such theories must transform
covariantly under Weyl transformations. This imposes additional restrictions
on the stress tensor at higher orders in the derivative expansion. In summary,
for a conformal fluid, the stress tensor up to first order has the form
\begin{equation}\label{ft}
T^{ab}= \alpha\,  T^{d} \left(d \, u^a u^b + \gamma^{ab} \right)
- \eta' \, T^{d-1} \sigma^{ab}
\end{equation}
where $\alpha$ and $\eta'$ are pure numbers and $\eta' \geq 0$. 

The constraints on allowed forms of the constitutive relations at higher order are more complicated. The most general allowed equations of second order fluid dynamics have largely (but perhaps not completely) been worked out in \cite{Romatschke:2009kr}. In the
following we will determine the second order fluid equations for ${\cal N}=4$
Yang Mills at strong coupling using the fluid/gravity duality.

\subsection{The Navier-Stokes scaling limit}
\label{s:nsscale}

An interesting fact about the equations of relativistic (or any other
compressible) fluid dynamics is that they reduce to a universal form under
a combined low amplitude and long wavelength scaling.
Consider a uniform fluid at rest, perturbed so that the amplitude in velocity fluctuations
is small (scales like $\epsilon$) and the amplitude in temperature
fluctuations is smaller (scales like $\epsilon^2$). We also require that
the wavelength of spatial fluctuations is large (scales like $1/\epsilon$)
and that their temporal scale even larger (scales like $1/\epsilon^2$).
We then take the strict $\epsilon \to 0$ limit. In this limit:
\begin{itemize}
\item[1)] The fluid is non-relativistic, as all velocities are parametrically
smaller than the speed of light.
\item[2)] The fluid is incompressible, as all velocities are parametrically
smaller than the speed of sound (recall that a sound wave is a compression
wave, and that fluid flows at velocities smaller than the speed of sound
are effectively incompressible).
\item[3)] The temporal component of the energy conservation equations reduces,
at leading order ${\cal O}(\epsilon^2)$ to the continuity equation
${\vec \nabla} \cdot {\vec v}=0$. We use the symbol ${\vec v}$ for the non-relativistic spatial velocity.
\item[4)]  The spatial component of the energy conservation equations
reduce at leading order, ${\cal O}(\epsilon^3)$, to the famous (non-relativistic) Navier-Stokes equations
\begin{equation}\label{ns}
{\dot {\vec v}} + {\vec v} \cdot { \vec \nabla } \, {\vec v}= -{\vec \nabla} P
+ \nu \, \nabla^2 {\vec v}
\end{equation}
with kinematic viscosity
$$\nu=\frac{\eta}{\rho_0 + P_0}.$$
where $\rho_0$ and $P_0$ are the background values of the density and
pressure of the fluid.
\end{itemize}

Note that the Navier-Stokes equations are homogeneous neither in amplitude of
fluctuations (the convective term is non-linear), nor in derivatives (the
viscous term is quadratic in derivatives). All terms retained in
\eqref{ns} are equally important in the $\epsilon \to 0$ limit; in particular, the parameter $\nu$ may be set to unity by a uniform rescaling of space and time. In contrast, the viscosities $\eta, \zeta$ give a subleading correction to ideal
fluid dynamics in the derivative expansion of relativistic fluid dynamics.

 Taking the spatial divergence of \eqref{ns} one sees  
that the pressure may be solved for in terms of the velocity field {\it on any
given time slice}; so the pressure is not an independent degree of freedom.
The initial data of the Navier-Stokes equations comprise just the components
of a divergence-free velocity field specified on any time slice. This reduction is not surprising
since the sound waves are being projected out in this limit.  

The incompressible Navier-Stokes equations \eqref{ns} describe a wide
variety of sometimes extremely complicated phenomena like turbulence.
Despite the fact that these equations have been studied for almost 200 years,
their non-linear phenomenology remains rather poorly understood.
One of the hopes of the fluid/gravity correspondence is to shed a new (geometric) light on some of these issues.

\section{Perturbative construction of gravity solutions}
\label{s:pertthy}

We have seen that fluid dynamics can be treated systematically as the theory of long wavelength fluctuations about thermal equilibrium. We are now going to construct gravitational solutions dual to fluid flows by formalizing this intuition to set up an algorithmic procedure to construct slowly varying dynamical black hole spacetimes as solutions to \eqref{Eeq}.

\subsection{Global thermal equilibrium from gravity}
\label{s:}
The starting point from the gravitational perspective is a solution that corresponds to global thermal equilibrium. For the moment let us consider a conformal field theory on Minkowski space $\gamma_{ab} = \eta_{ab}$.  From the gauge/gravity correspondence we know that the dual geometry in the bulk is the planar Schwarzschild-AdS$_{d+1}$ geometry, whose metric in static coordinates (with $x^a  = \{t,y^i\}$) is given by 
\begin{equation}
ds^2 = -r^2\, f(r/T) \, dt^2 + \frac{dr^2}{r^2\, f(r/T) }+ r^2\, \delta_{ij} \, dy^i\, dy^j \ , \qquad \ 
f(r) \equiv 1-\left(\frac{4\pi}{d\, r}\right)^d\!.
\label{sads1}
\end{equation}	
This is a one-parameter family of solutions, parameterized in terms of the black hole temperature $T$, which determines the horizon radius, $r_+ \equiv \frac{4 \pi \, T}{d}$, where $f$ vanishes. It is easy to generate a $d$-parameter family of solutions by boosting (\ref{sads1}) along the  translationally invariant spatial directions $y^i$, leading to a solution parameterized by a (normalized) timelike velocity field $u^a$.  The parameters which characterize the bulk solution are precisely the basic hydrodynamical degrees of freedom, viz., temperature $T$ and velocity $u^a$
of the black hole. It is easy to see that the solution induces on the Minkowski boundary of the AdS$_{d+1}$ spacetime a stress tensor which precisely takes the ideal fluid form of \eqref{Teq} with thermodynamic parameters specified by \eqref{cftPetc}. The normalization constant $\alpha$ is fixed by the gravitational theory to be $\alpha =  \frac{\pi^d}{16\pi\,G_N^{(d+1)}}$.

Consider this $d$-parameter family of boosted planar Schwarzschild-AdS$_{d+1}$ geometries, each of which holographically map to an ideal (conformal) fluid living on ${\bf R}^{d-1,1}$ (endowed with the Minkowski metric). This fluid is in global thermal equilibrium and one should be able to describe the long wavelength but arbitrary-amplitude fluctuations away from equilibrium via hydrodynamics. This class of fluctuations  has a bulk geometric avatar;  we  now describe an algorithmic procedure that enables us to construct such asymptotically locally AdS$_{d+1}$ black hole geometries which are  generically inhomogeneous and dynamical. This procedure crucially relies on the fact that hydrodynamics, as indeed any effective field theory, can be systematically studied in a gradient expansion.

\subsection{The perturbation theory}
\label{s:perthy}

We start by considering perturbations to  the {\it seed} geometry characterizing equilibrium. Take the boosted planar Schwarzschild-AdS$_{d+1}$ spacetime (\ref{sads1}), for convenience rewritten in ingoing coordinates so as to remove the coordinate singularity on the horizon, and replace the parameters $u_a$ and $T$ by functions of the boundary coordinates $x^a$, 
\begin{equation}   
ds^2 =-2\,u_{a}(x)\,dx^{a} \,dr -r^2\, f\left( r/T(x) \right) \,
u_{a}(x) \, u_{b} (x) \, dx^{a}\, dx^{b} +
r^2\, P_{ab}(x) \, dx^{a} \, dx^{b} \ ,
\label{boostedg0}
\end{equation}	
with $f(r)$ as specified in (\ref{sads1}) and $P_{ab}$ given by
(\ref{pdef}) with $\gamma_{ab}= \eta_{ab}$.
This metric, which we henceforth  denote as $g^{(0)}_{\mu\nu}(T(x), u^a(x))$, is not a solution to Einstein's equations. It however has two felicitous features: (i) it is regular for all $r>0$ and (ii) if the functions $T(x)$ and $u^a(x)$ are chosen so as to have small derivatives, then it can be approximated in local domains by a corresponding boosted black hole solution.
These observations lead us to consider an iterative procedure to correct (\ref{boostedg0}) order by order in a gradient expansion. 
We will find that we  however cannot specify just any slowly varying $T(x)$ and $u^a(x)$ (recall that $x^a$ included the temporal direction).
A true solution to Einstein's equations is obtained only when the functions $T(x)$ and $u^a(x)$ in addition to being slowly varying satisfy a set of equations which happen to be precisely the conservation equations of fluid dynamics.  Let us record that we have fixed gauge by setting $g_{rr}=0$ and $g_{ra}=-u_a$.

Since we want to keep track of the derivatives with respect to the boundary coordinates, it is useful to introduce a book-keeping parameter $\varepsilon$ and regard the variables of the problem as functions of rescaled boundary coordinates $\varepsilon x^a$. At the end of the day $\varepsilon$ may be set to unity. With this in mind, let us consider the corrections to the seed metric in a gradient expansion:
\begin{equation}
g_{\mu\nu} = \sum_{k=0}^\infty \, \varepsilon^k g^{(k)}_{\mu\nu}(T(\varepsilon x) ,u^a(\varepsilon x) ) \ , 
\ u^a = \sum_{k=0}^\infty \varepsilon^k u^{a\,(k)}(\varepsilon x)  \ , 
\  T = \sum_{k=0}^\infty \varepsilon^k  T^{(k)}(\varepsilon x) 
\label{gbuexpn}
\end{equation}	
where the correction pieces $g_{\mu\nu}^{(k)}$, $u^{a\,(k)}$ and $T^{(k)}$ are to be determined by solving Einstein's equations to the $k^{\rm th}$ order in the gradient expansion. The ansatz (\ref{gbuexpn}) should therefore be inserted into the Einstein's equations (\ref{Eeq}) and the result expanded in powers of $\varepsilon$. 

Let us examine the resulting structure in abstraction first. For the sake of argument, assume that we have determined $g_{\mu\nu}^{(m)}$ for  $m \leq n-1$ and $T^{(m)}$ and $u^{a\,(m)}$ for $m \leq n-2$. At order $\varepsilon^n$ one finds that Einstein's equations reduce to a set of {\em inhomogeneous linear differential equations} whose structure can be schematically written as
\begin{equation}
{\mathbb H} \left[g^{(0)}(T^{(0)}, u^{a\,(0)}) \right] \, g^{(n)} = \;s_n 
\label{homoop}
\end{equation}	
where we have dropped the spacetime indices for notational clarity
(c.f.,\ \cite{Bhattacharyya:2008jc,Rangamani:2009xk} for the explicit equations). Since each derivative with respect to $x^a$ is accompanied by a power of $\varepsilon$, it follows that  the linear operator ${\mathbb H}$ is constructed purely from the data of the equilibrium Schwarzschild-AdS$_{d+1}$ geometry. This means that ${\mathbb H}$ is at most a second-order differential operator with respect to the radial variable $r$. Moreover, it has to be independent of  $n$. Thus the perturbation theory in $\varepsilon$ is ultra-local in the boundary coordinates, implying that we can solve the equations of motion of the bulk spacetime point by point on the boundary!   

On the right hand side of (\ref{homoop}) we collect all order $\varepsilon^{n}$ terms which do not have explicit radial derivatives into a source term $s_n$, which is then a complicated construct involving contributions from different orders in perturbation theory. It is a local expression of $(n-m)^{\rm th}$ order in boundary derivatives of $T^{(m)}$ and $u^{a\,(m)}$ for $m \leq n-1$, and ascertaining it is the most substantial part of the computation.

The reader may be puzzled by the following aspect of \eqref{homoop}:  while we have $\frac{d(d+1)}{2}$ equations, we have only $\frac{d(d-1)}{2}$  variables after fixing the gauge redundancy. This implies that a subset of Einstein's equations has a distinguished status as constraint equations, while the remainder are the physical dynamical equations.

To understand this let us examine the differential equations (\ref{homoop}) by invoking the canonical split of our bulk coordinates $X^\mu = (r, x^a)$. The $E_{ra}$ equations are the momentum constraint equations for `evolution' in the radial direction. 
These equations are special in several ways. To start with, they need only 
be satisfied on a single $r$ slice; the `dynamical' equations ($E_{ab}$) then 
ensure that they will be solved on every $r$ slice. For this reason, it is 
consistent to study these equations just at the boundary, where they turn out
to reduce merely to the equations of conservation of the boundary stress 
tensor
\begin{equation}
\nabla_a T_{(n-1)}^{\; a b}=0  \ .
\label{cst}
\end{equation}	
(See \S\ref{s:sten2} for the definition of the boundary stress tensor.)
Note that at $n^{\rm th}$ order  the equations \eqref{cst} depend only on the 
boundary stress tensor built out of the spacetime metric at order $n-1$. This is because \eqref{cst} has an explicit boundary derivative which 
carries its own effective power of $\varepsilon$. The net upshot is that 
the unknown metric $g^{(n)}$ does not enter the equations \eqref{cst} at all 
(the operator ${\mathbb H}$ in \eqref{homoop} vanishes for these solutions). 
Hence, \eqref{cst} is instead a constraint on the solution already obtained 
at one lower order in perturbation theory. As we will see below, the solution
for $g^{(n)}$ of the dynamical equations at each order in perturbation theory 
is uniquely obtained in terms of the 
previous order solution, and so, ultimately, in terms of the velocity 
and temperature fields that enter the starting ansatz (the zeroth-order term 
in perturbation theory). Consequently, \eqref{cst} 
is an equation which constrains the starting velocity and temperature 
fields, and turns out to be the equation of boundary fluid dynamics. 

The remaining equations $E_{rr}$ (the `Hamiltonian constraint' for radial evolution) and $E_{ab}$ are dynamical equations with the operator ${\mathbb H}$ being a second order differential operator in $r$. Exploiting the spatial rotational symmetry of the seed solution, these equations can be decoupled and solved by quadratures,
\begin{equation}
g^{(n)} = {\rm particular} (s_n) + {\rm homogeneous}({\mathbb H}) \ .
\label{}
\end{equation}	
A unique solution to the dynamical equations
is obtained upon specification of boundary conditions: normalizability at infinity and regularity in the interior for all $r > 0$. These turn out to specify the solution completely\footnote{Modulo the fact that the operator ${\mathbb H}$ has zero modes which are to be accounted for by re-definitions of the background values of $T$ and $u^a$.} and  one ends up with a regular black hole geometry at each given order in the $\varepsilon$ expansion.

In summary, at any order in the perturbative expansion one solves the constraint
equations, enforcing fluid dynamical equations on the `initial' data. One then 
solves for the corrected metric. This correction feeds into the constraint 
equations giving corrected equations of fluid dynamics, and so on. The process
may be iterated to any desired order, thereby yielding 
a systematic derivative expansion of the equations of fluid dynamics.

\section{Results at 2nd order}
\label{s:secord}

Having seen abstractly the iterative procedure which perturbatively corrects the seed metric to obtain a solution to Einstein's equations at arbitrary order in the gradient expansion, we now turn to the results of this construction (for now still considering only energy-momentum transport on the boundary). While our discussion so far has been restricted to the case of a flat boundary metric $\gamma_{ab} = \eta_{ab}$, the observation we made about the ultra-locality of the perturbation theory allows us to immediately generalize to slowly-varying curved boundary metrics. Given a metric $\gamma_{ab}$ on the boundary, we can exploit the freedom to pass over to a Gaussian normal coordinate chart about the point under consideration, and account for the curvatures which arise starting with the second order in the $\varepsilon$ expansion via the computation of appropriate source terms. We will therefore present the results below for this more general setting. Before we do so, we will take the opportunity to review a beautiful technical framework developed by \cite{Loganayagam:2008is} to simplify the results for conformal fluids.

\subsection{Weyl-covariant formalism}
\label{s:weylc}

The vacuum AdS$_{d+1}$ spacetime is dual to the vacuum state of a conformal field theory. If we are interested in the hydrodynamic description of the latter on a background manifold ${\cal B}_d$, then rather than focusing on the metric $\gamma_{ab}$ of this geometry, we can consider the conformal class of metrics $\left({\cal B}_d, {\cal C}\right)$. On this conformal class there is a natural derivative operator, defined through a Weyl connection, which efficiently keeps track of Weyl transformation properties of various operators. This is all the more natural in the context of fluid dynamics where there is a distinguished vector field, the velocity $u^a$, defined to be the (normalized) timelike eigenvector of the stress tensor.

Let us first start with local Weyl rescalings of the boundary metric which transforms homogeneously, i.e.,
\begin{equation} 
 \gamma_{ab} = e^{2\phi}\; \widetilde{\gamma}_{ab} \qquad \Leftrightarrow \qquad
  \gamma^{ab} = e^{-2\phi}\;\widetilde{\gamma}^{ab} ,
\label{conftransf}
\end{equation}
We will call a tensor ${\cal Q}$ with components ${\cal Q}_{a_1 \cdots a_n}^{\ b_1 \cdots b_m}$ conformally covariant and of weight $w$ if it transforms homogeneously under Weyl rescalings of the metric, i.e., ${\cal Q} = e^{-w\,\phi} \, \widetilde{{\cal Q}}$ under (\ref{conftransf}).
The velocity field $u^a$ transforms as a weight $1$ tensor while the stress tensor $T^{ab}$ of a conformal fluid has weight $(d+2)$ in $d$-spacetime dimensions.

One defines a class of torsionless connections, called the Weyl connections, characterized by a connection one form ${\cal A}_a$,  whose associated covariant derivative $\nabla^{\text{Weyl}}$ captures the fact that the metric transforms homogeneously under conformal transformations (with weight $-2$). In particular, for every metric in the conformal class ${\cal C}$,
\begin{equation}
\nabla^{\rm Weyl}_a \gamma_{bc}  = 2 \, {\cal A}_a\, \gamma_{bc} \,.
\label{wcmet}
\end{equation}	
Given this derivative structure, we can go ahead and define a Weyl covariant derivative  ${\cal D}_a = \nabla^{\rm Weyl}_a + w \,  {\cal A}_a$
which is metric compatible and whose action on tensors transforming homogeneously with weight $w$  (i.e., ${\cal Q}^{a\cdots}_{b \cdots} = e^{-w\, \phi} \,  \widetilde{{\cal Q}}^{a\cdots}_{b \cdots}$)  is given by
\begin{eqnarray}
{\cal D}_c {\cal Q}^{a\cdots}_{b \cdots} &\equiv& \nabla_c {\cal Q}^{a\cdots}_{b \cdots} + w\, {\cal A}_c\,{\cal Q}^{a\cdots}_{b \cdots}\nonumber \\
&&  +\left( \gamma_{c d} \, {\cal A}^a - \delta^a_c \, {\cal A}_d - \delta^a_d \,{\cal A}_c\right) \,  {\cal Q}^{d\cdots}_{b \cdots} + \cdots \nonumber \\
&&- \left( \gamma_{cb}\, {\cal A}^d - \delta^d_c \, {\cal A}_b - \delta^d_b \, {\cal A}_c\right) \, {\cal Q}^{a\cdots}_{d \cdots} -\cdots \ .
\label{weylcd}
\end{eqnarray}	
The connection has been defined so that the Weyl covariant derivative of a conformally covariant tensor transforms homogeneously with the same weight as the tensor itself.

In hydrodynamics we will require that the Weyl covariant derivative of the fluid velocity be transverse and traceless, 
\begin{equation}
u^a \, {\cal D}_a u^b = 0 \ , \qquad {\cal D}_a u^a = 0 \ ,
\label{}
\end{equation}	
which enables one to uniquely determine the connection one-form ${\cal A}_a$ to be  the distinguished vector field 
\begin{equation}
{\cal A}_a= u^c\, \nabla_{\! c} \, u_a - \frac{1}{d-1}  \, u_a 
\, \nabla_{\!c} \, u^c = a_a - \frac{1}{d-1}\, \theta \, u_a  \  , 
\label{cadef}
\end{equation}	
built from the fluid velocity field. 

One can rewrite the various quantities appearing in the gradient expansion of the stress 
tensor in this Weyl covariant notation. For instance, at first order in derivatives, we have the shear and vorticity constructed from the velocity field:
\begin{equation}
\sigma^{ab} = {\cal D}^{(a} u^{b)} \ , \qquad 
\omega^{ab} = -{\cal D}^{[a} u^{b]} \ ,
\label{udecdefs2}
\end{equation}
both of which have weight $w =3$.  The fluid dynamical equations, viz., stress tensor conservation, are simply ${\cal D}_a \, T^{ab} = 0$ in this Weyl covariant language (which is equivalent to  (\ref{stc}) since (\ref{weylcd}) with $w=d+2$ gives
${\cal D}_a \, T^{ab} = \nabla_{\!a}\, T^{ab} + T_a^{\ a} \, {\cal A}^b$ and the conformal fluid stress tensor must be traceless).

\subsection{Generic asymptotically AdS black hole metric}
\label{s:met2}
We now have at our disposal all the technical machinery necessary to present the results for the gravity dual of non-linear fluid dynamics. By a suitable choice of gauge (a slight generalization of the Eddington-Finkelstein coordinates), one can express the bulk metric $g_{\mu\nu}$ in the form
\begin{equation} 
ds^2 = - 2 \,   u_a(x) \, dx^a \,\left( dr + {\mathfrak V}_b(r,x)\,\,dx^b\right)+ {\mathfrak G}_{ab}(r,x) \, dx^a\, dx^b \ , 
\label{formmetw}
\end{equation}
where the fields ${\mathfrak V}_a$ and ${\mathfrak G}_{ab}$ are functions of $r$ and $x^a$ which admit an expansion in the boundary derivatives. In the parameterization used in \cite{Bhattacharyya:2008mz} one finds the metric functions are given up to second order in derivatives as:
\begin{equation} 
\begin{aligned}
{\mathfrak V}_a & = r\, {\cal A}_a - {\cal S}_{ac}\,u^c - {\mathfrak v}_1(r/T)\, P^{\; b}_a\, {\cal D}_c \, \sigma^{c}_{\ b} \\
&\qquad  +u_a \, \left[\frac{1}{2}\, r^2 \, f(r/T) + \frac{1}{4}\, \left(1-f(r/T)\right) \, \omega_{cd}\, \omega^{cd} + {\mathfrak v}_2(r/T) \, \frac{\sigma_{cd}\, \sigma^{cd}}{d-1}\right]  \\
{\mathfrak G}_{ab} &= r^2\, P_{ab}  - \omega_{a}^{\ c}\,\omega_{cb}+ 2\, (r/T)^2\, {\mathfrak g}_1(b\,r)\, \left[\frac{4\pi T}{d}\, \sigma_{ab} + {\mathfrak g}_1(r/T) \, \sigma_a^{\ c}\,\sigma_{cb} \right] \\
&\qquad - {\mathfrak g}_2(b\,r)  \,\frac{\sigma_{cd}\, \sigma^{cd}}{d-1} \, P_{ab} 
- {\mathfrak g}_3(r/T) \, \left[{\mathfrak T}_{1ab} + \frac{1}{2}\, {\mathfrak T}_{3ab}   + 2\,{\mathfrak T}_{2ab} \right] \\
&\qquad + {\mathfrak g}_4(r/T)\,  \left[{\mathfrak T}_{1ab} +  {\mathfrak T}_{4ab}   \right] .
\end{aligned}
\label{met2w}
\end{equation}
Here ${\cal S}_{ab} = \frac{1}{d-2}\left({\cal R}_{ab}-\frac{{\cal R} \, }{2(d-1)}\, \gamma_{ab}\right)$ is the  Schouten tensor of the boundary metric, where
the Weyl covariant curvature tensors are 
\begin{equation}
\begin{split}
{\cal R}_{ab} &= R_{ab} -(d-2)\left(\nabla_{\! a} \,  {\cal A}_b + {\cal A}_a \, {\cal A}_b -{\cal A}^2 \, \gamma_{ab}  \right)-g_{ab}\nabla_{\! c} \, {\cal A}^c - {\cal F}_{ab}\\
{\cal R} & \equiv {\cal R}_{a}^{\ a} = R -2\, (d-1) \, \nabla_{\! c} \, {\cal A}^c + (d-2)(d-1) \, {\cal A}^2 \ 
\end{split}
\label{ricEin:eq}
\end{equation}
with ${\cal F}_{ab} \equiv \nabla_{\! a} \,  {\cal A}_b  - \nabla_{\! b} \,  {\cal A}_a $.
Apart from the shear and vorticity tensors (\ref{udecdefs2}) constructed from the fluid velocity, we also encounter four of the five second order tensors which form a Weyl covariant basis,
\begin{equation}
\begin{aligned}
&{\mathfrak T}_1^{ab} =2\, u^c \, {\cal D}_c \sigma^{ab} \ , \quad 
 {\mathfrak T}_2^{ab} = C^{acbd}\,u_c \,u_d \ ,  \\
 &{\mathfrak T}_3^{ab}  =4\,\sigma^{c\langle a}\, \sigma^{b\rangle}_{\ c}  \ , \quad 
  {\mathfrak T}_4^{ab} =  2\, \sigma ^{c\langle a}\, \omega^{b\rangle}_{\ c}  \ , \quad  {\mathfrak T}_5^{ab}=   \omega ^{c\langle a}\, \omega ^{b\rangle}_{\ c} \ .
\end{aligned}
\label{winv2der}
\end{equation}	
Note that the tensor ${\mathfrak G}_{ab}$ is clearly transverse, since it is built out of operators that are orthogonal to the velocity, and it can be inverted via the relation 
$\left({\mathfrak G}^{-1}\right)^{ac} \, {\mathfrak G}_{cb}  = P^a_{\ b}\, $. The induced metric on the boundary in these coordinates takes the form:
\begin{equation}
\gamma_{ab} = \lim_{r\to \infty} \, \frac{1}{r^2} \, \left({\mathfrak G}_{ab} -2\, u_{(a}\, {\mathfrak V}_{b)} \right) ,
\label{bdymet2w}
\end{equation}	
which is crucially used to raise and lower the boundary indices. 

Finally, the various functions ${\mathfrak g}_{i}$ and ${\mathfrak v}_{i}$ appearing in the metric are given in terms of definite integrals once one has inverted the operator ${\mathbb H}$:  
\begin{equation}
\begin{aligned}
{\mathfrak g}_1(y) &= \int_y^\infty\, d\zeta\, \frac{\zeta^{d-1}-1}{\zeta\, \left(\zeta^d -1\right)} \\
{\mathfrak g}_2(y) &= 2\,y^2 \, \int_y^\infty \frac{d\xi}{\xi^2} \int_\xi^\infty\, d\zeta\,\zeta^2 \, {\mathfrak g}_1'(\zeta)^2 \\
{\mathfrak g}_3(y) & =y^2 \, \int_y^\infty\, d\xi\, \frac{\xi^{d-2}-1}{\xi \, \left(\xi^d -1\right)} \\
{\mathfrak g}_4(y)& = y^2 \, \int_y^\infty\,  \frac{d\xi}{\xi \, \left(\xi^d -1\right)}
\int_1^\xi\,d\zeta\, \zeta^{d-3}\bigg(1+ (d-1)\,\zeta\,{\mathfrak g}_1(\zeta) + 2\, \zeta^2\,{\mathfrak g}'_1(\zeta)\bigg) \\
{\mathfrak v}_1(y) &=\frac{2}{y^{d-2}} \, \int_y^\infty \, d\xi \; \xi^{d-1}\int_\xi^\infty\, d\zeta\, \frac{\zeta-1}{\zeta^3\, \left(\zeta^d-1\right)}\\ 
{\mathfrak v}_2(y) &= \frac{1}{2\, y^{d-2}} \, \int_y^\infty\; \frac{d\xi}{\xi^2}\,\bigg[1-\xi\,(\xi-1) \,{\mathfrak g}'_1(\xi)  -2 \,(d-1)\,\xi^{d-1} \\
&\qquad + \left(2\, (d-1)\,\xi^d - (d-2)\right) \, \int_\xi^\infty\, d\zeta\, \zeta^2\,{\mathfrak g}'_1(\zeta)^2 \bigg]  .
\end{aligned}
\label{fmetfnsw}
\end{equation}
The asymptotic behavior of these functions ${\mathfrak g}_i(r/T)$ and ${\mathfrak v}_i(r/T)$ is important for the stress tensor computation of \S\ref{s:sten2} and can be found in \cite{Bhattacharyya:2008mz}.

\subsection{Event horizon and entropy current}
\label{s:eventhor}

The metric  (\ref{formmetw}), (\ref{met2w}) solves Einstein's equations to second order in the gradient expansion, provided the first order stress tensor
(which takes the form (\ref{ft}) with the coefficients extracted from the 1st order bulk metric, and given explicitly below in \S\ref{s:sten2}) satisfies the hydrodynamic conservation equations. While this already establishes a firm connection between solutions of Einstein's equations and those of fluid dynamics (in one lower dimension), it is imperative to establish that the bulk geometry we describe is regular everywhere outside the curvature singularity at $r=0$. 

Although one can utilize the behavior of the metric functions and iteratively argue that the sources are regular order by order in perturbation theory, it is convenient to establish once and for all that what one has constructed is a black hole spacetime with a regular event horizon. Doing so involves ascertaining the location of the event horizon.  A-priori, this sounds like a tall order, especially given that explicit solution is contingent on having solved the fluid equations. Moreover, as is well known, the event horizon is a teleological concept (it is the boundary of the past of future null infinity) whose determination requires knowing the entire future history of the spacetime. However, with one key assumption of late-time relaxation which is natural from fluid dynamics, it turns out to be possible to determine the location of the event horizon {\it locally} within our gradient expansion.  Apart from showing regularity, this has the additional virtue of enabling us determine a natural entropy current for fluid dynamics \cite{Bhattacharyya:2008xc}.

Since generic flows of dissipative fluids tend to approach global equilibrium  at late times, it follows that the corresponding event horizon has to approach the radial position determined by the local late-time temperature of the fluid. In particular, we look for a null co-dimension one surface given by the equation $S_{\cal H}(r,x) = r - r_{\cal H}(x)=0$ with the correct asymptotics. 
The function $r_{\cal H}(x)$ should be parameterized within the gradient expansion $r_{\cal H}(x) = \frac{4 \pi \, T(x)}{d} + \sum_k \, \varepsilon^k \, r_{(k)}(x)$. The corrections $r_{(k)}(x)$ are determined by solving the null condition $g^{\mu\nu} \, \partial_\mu S_{\cal H} \,  \partial_\nu S_{\cal H} =0$. The resulting equations are algebraic for $r_{(k)}$ and to second order in gradients one finds that, for the solution (\ref{formmetw})-(\ref{fmetfnsw}),

\begin{equation}
r_{\cal H}(x) = \frac{4 \pi \, T(x)}{d}+  \frac{d}{4 \pi \, T(x)}\, \left(\aleph_1 \, \sigma_{ab}\, \sigma^{ab} + \aleph_2 \, \omega_{ab}\, \omega^{ab} + \aleph_3\, {\cal R}\right)
\label{}
\end{equation}	
with
\begin{equation}
\begin{aligned}
\aleph_1 &= \frac{2\,(d^2 +d-4)}{d^2\,(d-1)\,(d-2)} - \frac{2 \,{\mathfrak v}_2(1)}{d\,(d-1)}\\
\aleph_2 &= -\frac{d+2}{2\,d\,(d-2)} \ , \qquad \aleph_3  = -\frac{1}{d\,(d-1)\,(d-2)} \ .
\end{aligned}
\label{alephdef}
\end{equation}	
This indeed establishes that the solutions we have constructed in \S\ref{s:met2} to 2nd order qualify to be called inhomogeneous, dynamical black holes.

Note that in general, beyond the leading order, the horizon position and generators  are not simply given by the corresponding fluid temperature and    velocity (for example, while the horizon generators must be vorticity-free,  $\omega_{ab}$ need not vanish for the boundary fluid).  In some sense, while the black hole horizon is distinguished in the bulk, physics appears simpler when expressed in terms of the fluid data living on the boundary.

Having determined the event horizon of the gravity solution, we immediately have access to an important hydrodynamic quantity, viz., the {\em entropy current}. For a black hole spacetime it is natural to view the area of the event horizon as an entropy {\it a la} Bekenstein-Hawking \cite{Bekenstein:1973ur,Hawking:1974sw}. In fact, by suitably foliating the event horizon with spatial slices (propagated forward by the null generator), we can equivalently talk about an area $(d-1)$-form $a_{\cal H}$ on these slices. Since we imagine the dual fluid living on the boundary of the spacetime, it is natural to pull-back this area form out to the boundary. A canonical choice is to pull-back along  radially ingoing null geodesics \cite{Bhattacharyya:2008xc}, which is quite easy to implement for the metric (\ref{formmetw}), where the lines of $x^a = {\rm constant}$ are precisely such geodesics. We then have a $(d-1)$-form on the boundary which can be dualized to a one-form or equivalently a current $J^a_s$, which is the entropy current on the boundary. Not only does this definition agree with the equilibrium notion of entropy of the fluid, but also thanks to the area theorem of black hole horizons, we are immediately guaranteed that this current has manifest non-negative divergence as demanded by the second law. The hydrodynamic entropy current takes the general form
\begin{equation}
\begin{split}
\, J^a_s &= s\,u^a +  \frac{s\, d^2}{(4 \pi \, T)^2} \, u^a \,\left(A_1 \,\sigma_{cd}\,\sigma^{cd}+A_2 \,\omega_{cd}\,\omega^{cd} +A_3 \,\mathcal{R}\,\right)\\
&\quad  +  \frac{s\, d^2}{(4 \pi \, T)^2} \,\left( B_1 \,{\cal D}_c \, \sigma^{ac} + B_2 \,{\cal D}_c \, \omega^{ac} \right) + \cdots
\end{split}
\label{ecurw}
\end{equation}
where $s$ is the entropy density and $A_{1,2,3}$, $B_{1,2}$ are a-priori arbitrary numerical  coefficients. While ${\cal D}_a J^a_s \geq 0$ only demands that $B_1 +2 \,A_3 =0$, the gravity solution (\ref{formmetw}) fixes all the coefficients in (\ref{ecurw}) explicitly. 
In particular, we obtain
\begin{equation}
\begin{split}
&s = \frac{1}{4\,G_N^{(d+1)}} \left(\frac{4\pi\, T}{d}\right)^{\! \! d-1} \  , \qquad  
A_1 = \frac{2}{d^2}\, (d+2) -\left( \frac{1}{2}\, {\mathfrak g}_2(1) +\frac{ 2}{d}\, {\mathfrak v}_2(1)\right) ,\\
&A_2 = -\frac{1}{2\,d} \ , \qquad B_1 = -2 \, A_3 = \frac{2}{d\, (d-2)} \ , \qquad B_2 = \frac{1}{d-2}  \ .
\end{split}
\label{sdengr}
\end{equation}	

We should note here that there is an ambiguity in pulling back the area-form from the event horizon to the boundary, for one can supplement the pull-back map with a boundary diffeomorphism, which affects the coefficient $A_1$ above. Since this just relabels boundary points in the gradient expansion, one is tempted to think of this ambiguity as unphysical. However, it is rather curious that if one tries to pull-back the area form from quasi-local horizons one encounters a shifted value of $A_1$ \cite{Booth:2011qy}, which suggests that there is perhaps more to this ambiguity than meets the eye.

\subsection{Stress tensor of dissipative fluid}
\label{s:sten2}

Given an asymptotically locally AdS$_{d+1}$ metric, one can construct a quasi-local boundary tensor which is manifestly conserved and is associated with the stress-energy-momentum tensor of the conformal field theory \cite{Henningson:1998gx, Balasubramanian:1999re}. To perform the computation one regulates the bulk spacetime by introducing an explicit cut-off at $r = r_\infty$. The boundary stress tensor is given in terms of the extrinsic curvature $K_{ab}$ of this surface, defined in terms of its unit outward pointing normal $n^a$ as  $
K_{ab} = \gamma_{ac}\, \nabla^c n_b$. In addition to the extrinsic curvature one also has contributions from the counter-terms necessary to obtain a finite boundary stress tensor. Denoting the curvatures of the boundary metric by $^\gamma \!R $ etc., this is given (to 2nd order) as
\begin{equation}
T_{ab} = \lim_{r_\infty \to \infty}\; \frac{-\,r_\infty^{d}}{8\pi \, G_N^{(d+1)}} \, \left[ K_{ab} - K \, \gamma_{ab} + (d-1)\, \gamma_{ab} - \frac{1}{d-2}\,  \left(  ^{\gamma}\!R_{ab} -\frac{1}{2}\, \, ^{\gamma}\!R \, \gamma_{ab}\right)\right] 
\label{BYstress}
\end{equation}	

For the gravity duals to fluid dynamics constructed in \S\ref{s:met2}, one finds that the boundary stress tensor takes the form \eqref{const} with the  dissipative part, at the first and second order, given by this gravitational construction to be
\begin{eqnarray}
\Pi_{(1)}^{ab} &=& -2\, \eta\,\sigma^{ab} \nonumber \\
\Pi_{(2)}^{ab} &=&  \tau_\pi\,\eta\, {\mathfrak T}_1^{ab} + \kappa\,{\mathfrak T}_2^{ab} + \lambda_1\, {\mathfrak T}_3^{ab} + \lambda_2\, {\mathfrak T}_4^{ab} + \lambda_3\, {\mathfrak T}_5^{ab} \  .
\label{fld2}
\end{eqnarray}	
where the tensors ${\mathfrak T}_i^{ab}$ were defined earlier in (\ref{winv2der}). With the tensor structure determined, one is just left with fixing the six transport coefficients, $\eta$, $\tau_\pi$, $\kappa$, and $\lambda_i$ for $i = \{1,2, 3\}$, which completely characterizes the flow of a non-linear viscous fluid with a gravitational dual. The transport coefficients for conformal fluids in $d$-dimensional boundary turn out to be 
\begin{equation}
\begin{aligned}
&\eta = \frac{1}{16 \pi\, G_N^{(d+1)}} \, \left(\frac{4\pi}{d}\,   T\right)^{d-1} 
\ , \\
&\tau_\pi = \frac{d}{4\pi\, T} \, \left[1 + \frac{1}{d}\, {\rm Harmonic}\left(\frac{2}{d} -1\right) \right] , \qquad \kappa = \frac{d}{2\pi\, (d- 2)} \,\frac{\eta}{T} \ , \\
& \lambda_1 = \frac{d}{8\pi}\, \frac{\eta}{T} \ , \qquad
\lambda_2 = \frac{1}{2\pi}\,  \text{Harmonic}\left(\frac{2}{d} -1\right) \, \frac{\eta}{T}\ , \qquad \lambda_3=0 \ . 
\end{aligned} 
\label{transpgend}
\end{equation}	
where ${\rm Harmonic}(x)$ is the harmonic number function. Setting $d=4$ in the above expressions and using the fact that $\text{Harmonic}(-\frac{1}{2}) = -2 \,\log(2)$ together with the replacement $\frac{1}{16\pi \, G_N^{(5)}} = \frac{1}{8\pi^2} \, N^2$, one can obtain the transport coefficients for $SU(N)$ ${\cal N}=4$ Super Yang-Mills theory \cite{Baier:2007ix,Bhattacharyya:2008jc}. This has been used for real data analysis from e.g.\ RHIC.

One immediate consequence of (\ref{transpgend}) and (\ref{sdengr}) is that our fluid saturates the famous bound on the viscosity to entropy density ratio, $\frac{\eta}{s} \ge \frac{1}{4\pi}$, \cite{Kovtun:2004de}. This bound is saturated by a large class of two-derivative theories of gravity, and it is indeed experimentally satisfied by all presently-known systems in nature. Intriguingly, cold atoms at unitarity and quark-gluon plasma both come near to saturating this bound \cite{Schafer:2009dj}. Its status in more general theories is currently under active debate \cite{Buchel:2008vz}.

Moreover, (\ref{transpgend}) reveals further intriguing relations between the coefficients, which hint at the specific nature of any conformal fluid which admits a gravitational dual.
For example, the result that $\lambda_3 = 0$ is universal but non-trivial from the fluid standpoint.  We also see that  $2 \, \eta \, \tau_\pi = 4 \lambda_1 + \lambda_2$   for all $d$; this in fact was shown to hold quite generally in a large class of two-derivative theories of gravity (including matter couplings)  \cite{Haack:2008cp}.

\section{Specific fluid flows and their gravitational analog}
\label{s:applications}

The construction presented above can be generalized in many interesting ways; however before indicating the most important of these in \S\ref{s:extensions}, we first pause to discuss some of the special cases of the framework explained in the preceding section. One of the reasons to discuss these special cases is that while we have demonstrated the existence of a map from the equations governing fluid dynamics to those governing the dynamics of gravity, we did not at any stage  solve the fluid equations explicitly. The felicitous feature of our construction was the ultra-locality along the boundary directions which allowed us to implement the construction in terms of local solutions to the conservation equations. Construction of novel fluid flows and generic behavior of the relativistic conservation equations are interesting (and perhaps hard) questions. Nevertheless there are some corners  where we can gain analytic control which serves not only as a check that the fluid/gravity solution set is non-empty, but also provides a point of contact with previous studies of the hydrodynamic regime in the AdS/CFT literature.

\subsection{Linearized setting: quasinormal modes}
\label{s:}

Above we have established a map between any solution of the equations of 
fluid dynamics and long wavelength solutions of Einstein gravity with a
negative cosmological constant. In order to find explicit 
gravitational solutions we need a class of explicit solutions to the 
equations of fluid dynamics. In this subsection and the next we will 
study such examples. 

It is of course easy to solve the equations of fluid dynamics, derived above,  
when linearized about static equilibrium. Utilizing translational invariance,
we search for solutions of the form 
$$u^a= \delta^a_t + \delta^a_j \, \delta v^j \, e^{i (\omega t + k^i y_i)} \,, \qquad T= T_0 + \delta T \,
e^{i (\omega t + k^i y_i)} \,,$$
with purely spatial velocity fluctuations $\delta v^j$. The resulting linear equations require that the matrix of coefficients $M(\omega, k)$ annihilate the length-$d$ column vector with entries $\delta v^i$ and $\delta T$. So the spectrum $\omega(k)$ is obtained as the roots of the $d^{\rm th}$ order polynomial $\text{det}(M)= 0$.
 At leading (ideal fluid) order, the $d$ 
roots to this equation turn out to be $\omega = \pm \frac{k}{\sqrt{d-1}}$ 
(the sound modes of the fluid) and  $\omega=0$ with degeneracy $d-2$ 
(the shear modes of the fluid). These modes and their corresponding
eigenvectors receive corrections at higher orders in the derivative 
expansion; in particular the shear modes pick up nontrivial $k$ 
dependence, $\omega \propto i\,k^2$, at first order. The explicit 
solutions are easily determined (see \cite{Bhattacharyya:2008jc}).  
Employing the fluid/gravity map then yields explicit linearized 
solutions of the Einstein equations \eqref{Eeq}  about the planar black hole background, whose study in fact predates
the fluid/gravity map by almost 10 years.

The spectral problem of linearized fluctuations about a black hole solution is of course a well-studied topic (cf., \cite{Berti:2009kk}). It is known that due to the presence of a horizon, black holes admit no 
normal modes. Instead, by imposing regularity at the future event horizon, one finds quasinormal modes, i.e., modes which have complex frequencies characterizing decay of perturbations at late times. Mathematically these are related to the poles of the retarded Green's function computed in the black hole background.

In terms of the gauge/gravity perspective, such quasinormal modes describe the timescale for return to thermal  equilibrium in the field theory \cite{Horowitz:1999jd}. Asymptotically AdS black holes host an infinite family of 
quasinormal modes. 
All except $d$ of these are `massive';
their frequency remains finite (and has a finite imaginary part or decay 
rate $\propto T$) even in the limit $k \to 0$.  However, planar black holes admit exactly $d$ special `massless' quasinormal
modes. These so-called hydrodynamic modes  can have arbitrarily low 
frequency at long spatial wavelengths and therefore fall within the 
long wavelength regime. These are precisely the 
sound and shear modes described above.

The fact that the dispersion relations of the massless quasinormal modes agrees
 with hydrodynamic dispersion relations was first demonstrated in the 
pioneering works \cite{Policastro:2002se,Policastro:2002tn}, who mapped the Schwarzschild-AdS quasinormal modes to sound and shear modes of the dual field theory (perturbed from thermal equilibrium).

\subsection{Rotating black holes in global AdS space}
\label{s:}

A second class of examples which accord analytic control are explicit solutions corresponding to stationary configurations in hydrodynamics. While on flat space there are no interesting stationary flows other than a uniformly boosted fluid  (whose dual is the seed solution we started with), it turns out that by a suitable choice of background geometry one can derive nontrivial flows. We now describe one such flow on the Einstein Static Universe (${\mathbb R} \times {\bf S}^{d-1}$) which allows one to make contact with rotating black holes in asymptotically (globally) AdS spacetimes.

Fluid flows of a conformal relativistic $d$-dimensional fluid on spatial ${\bf S}^{d-1}$ conserve angular momentum in addition to energy. The angular momentum on ${\bf S}^{d-1}$ is a rank-$d$ antisymmetric matrix, which can be brought to canonical form by an $SO(d)$
similarity transform (a rotation) and is therefore labeled by its $[d/2]$ inequivalent
eigenvalues. For every physically allowed choice of these $[d/2]$ angular momenta
together with the energy, an arbitrary fluid flow eventually settles down into an equilibrium stationary configuration.

The stationary configurations of viscous conformal fluids on spheres
turn out to be extremely simple. The velocity field is simply that of
a rigid rotation. Focusing on the case of $d=2n$ for concreteness,
the metric of a unit ${\bf S}^{2n-1}$ may be written in terms of the direction cosines $\mu_i$ as
\begin{equation}\label{metric}
ds_{{\bf S}^{2n-1}}^2=\sum_{i=1}^n \mu_i^2 \, d \phi_i^2 + d\mu_i^2 \ ,
\qquad {\rm where} \qquad \sum_{i=1}^n \mu_i^2=1  \,.
\end{equation}
In these coordinates the velocity  and temperature
fields of stationary flows take the form
\begin{equation}\label{vf}
u_s^a \, \partial_a= \gamma \left( \partial_t + \sum_{i=1}^n \omega_i \, 
\partial_{\phi_i} \right) \,, \quad 
T_s =  \gamma\, T_0\ , \quad \gamma = \left(1-\sum_{i=1}^m \omega_i^2 \mu_i^2\right)^{\! \! -\frac{1}{2}} 
\end{equation}
This flow is Weyl equivalent to a uniform velocity and temperature configuration on a confomally rescaled spacetime \cite{Bhattacharyya:2007vs}, i.e.,
\begin{equation}\label{vfw}
ds^2=\gamma^{2}\, \left(-dt^2+ ds_{{\bf S}^{2n-1}}^2\right) , \quad
u_s^a \, \partial_a = \partial_t + \sum_{i=1}^n \omega_i
\partial_{\phi_i} \,, \quad
T_s = T_0 \ .
\end{equation}

It is not difficult to verify that  \eqref{vfw} provides a non-dissipative solution to the equations of fluid dynamics at least
to second order (the solution is non-dissipative because it is stationary;
equivalently the divergence of the entropy current vanishes). The constant
parameters $\omega_i$ and $T_0$ of this solution turn out to have
thermodynamical significance: they are simply the angular velocity (chemical
potential for angular momentum) and temperature of the fluid configuration.

According to the AdS/CFT correspondence, the dual description of a
conformal field theory on ${\mathbb R} \times {\bf S}^{d-1}$  is simply asymptotically
global AdS$_{d+1}$ space. If we pump a large amount of energy and angular
momentum into global AdS$_{d+1}$ space and let the system relax, we expect
the eventual equilibrium configuration to be that of a large rotating black
hole. This reasoning leads to a prediction: the fluid/gravity map applied
to \eqref{vfw} should produce the (independently known) metric of large
rotating AdS$_{d+1}$ black hole, expanded to second order in the derivative
expansion. This prediction\footnote{
The first observation that the properties of large rotating black holes
should be reproduced from fluid dynamics was made in
\cite{Bhattacharyya:2007vs}, a precursor to the fluid/gravity map.
Rotating black holes were further studied in \cite{Bhattacharyya:2008ji}
in 4 dimensions, and fully analyzed in $d$ dimensions in
\cite{Bhattacharyya:2008mz}.  Here for illustration we give the
general result of \cite{Bhattacharyya:2008mz}, who find the explicit
coordinate transformation to rewrite the rotating black hole solution
in AdS$_{d+1}$ of \cite{Gibbons:2004uw} in fluid variables.} has been verified in detail
in the following manner. It is possible to transform the exactly known
metric of the rotating AdS$_{d+1}$ black holes to the fluid/gravity gauge described in
\S\ref{s:pertthy} above. This maneuver in fact turns out
to greatly simplify the rotating black hole metric which takes the
form (\ref{formmetw}) with
\begin{eqnarray} \label{bhm}
{\mathfrak V}_a(r,x) & =& r\, {\cal A}_a - {\cal S}_{ac}\,u^c
- \frac{(4 \pi \, T)^d \, r^2 \, u_a}{2 \, d^2 \,  \det \left[ r \,
\delta^{b}_{\ c} - \omega^{b}_{\ c}  \right]} \ , \nonumber \\
{\mathfrak G}_{ab}(r,x) & =&  r^2\, P_{ab}  - \omega_{a}^{\ c}\,\omega_{cb} \ .
\label{KerrAdSmet}
\end{eqnarray}
The above is an exact rewriting of the rotating AdS black holes (in even dimensions) of \cite{Gibbons:2004uw}.
The metric \eqref{bhm} may be expanded in derivatives simply
by expanding the inverse determinant in powers of $\omega$. By truncating this expansion at second order we recover exactly the metric dual to the
fluid flow \eqref{vfw}. It is rather remarkable the full black hole solution can be written economically within the fluid/gravity metric ansatz, perhaps suggesting greater utility for metrics of the form \eqref{formmetw}.

\subsection{Non-relativistic fluids}
\label{s:nonrelfg}

While relativistic fluids are interesting in  astrophysical or high energy plasma physics contexts, most fluids we encounter in everyday situations are non-relativistic. Furthermore, for many practical applications one is usually  interested in their dynamics in the incompressible regime, which is attained by projecting out the sound mode. It is natural to ask whether this regime is accessible to fluid/gravity; an affirmative answer is suggested by the discussion in \S\ref{s:nsscale}, namely one only needs to implement the Navier-Stokes scaling limit directly in the fluid/gravity solutions. This procedure has been carried out in \cite{Bhattacharyya:2008kq} to obtain the gravitational dual of non-relativistic incompressible fluid flows. In principle this provides a geometric window to explore phenomenologically interesting fluid flows.

\section{Extensions beyond conformal fluids}
\label{s:extensions}

The fluid/gravity correspondence was originally derived for the case of conformal fluids which are related to gravitational dynamics in an asymptotically AdS spacetime. Conformal
theories are rather special and `ordinary' fluids one encounters everyday deviate significantly from this behavior. Hence it would be useful to look for extensions of the basic framework which allows for these generalizations. As already indicated earlier, this can be done at the expense of complicating the system of gravitational equations, and accompanying loss of universality. Nevertheless, forays into these areas have revealed very interesting lessons about fluid dynamics in general that transcend the fluid/gravity correspondence. In this section we take stock of some of these developments.

\subsection{Non-conformal fluids}
\label{s:nonconf}

The first generalization we consider is provided by a handy trick to obtain a particular class of non-conformal theories.
It turns out that by exploiting the gauge/gravity duality for a special class of theories, viz.,  theories that naturally arise on the world-volume of D$p$-branes, one finds a surprisingly tractable class of non-conformal fluids. 

Let us first consider the special case of the D4-brane which is a solution of the equations of 10-dimensional IIA supergravity. IIA  supergravity is the dimensional reduction of 11-dimensional supergravity  on ${\bf S}^1$, and the D4-brane solution is the dimensional reduction of a M5 brane solution that wraps the ${\bf S}^1$.  The near horizon 
geometry of the M5 brane solution is AdS$_7 \times {\bf S}^4$, and the 11-dimensional equations admit a consistent truncation to 7-dimensional equations of motion \eqref{Eeq} involving gravitational dynamics with a negative cosmological constant. Compactifying further on an ${\bf S}^1$ and restricting to the zero momentum sector on this circle, yields a further consistent truncation of this 
seven dimensional set of equations. The  resulting 6-dimensional equations are simply the Einstein-dilaton equations about the D4-brane background. It follows that the Einstein-dilaton 
equations constitute a  consistent truncation of the equations of IIA 
supergravity about the D4 near-horizon background (as is easily independently
verified).

It turns out that the fluid dynamics dual to the 
long wavelength fluctuations about the thermal M5 brane solution is simply
that computed in \S\ref{s:pertthy}, for the special case $d=6$.
We want to focus on gravitational solutions corresponding to fluid flows independent of one of the boundary directions $x^i$ (the direction of the ${\bf S}^1$ in the paragraph above). These lie within the six dimensional consistent truncation  described above. But these are simply the gravitational solutions dual  to fluid flows on the world volume theory of the D4-brane. In other words,  the fluid dynamics of the D4-brane world volume theory is a dimensional 
reduction of the conformal fluid dynamics on the world volume of the M5 brane. 
Moreover the gravitational duals to D4-brane fluid flows are very easily 
obtained from the KK reduction of the results of \S\ref{s:pertthy}.
Note that the dimensional reduction of conformal fluid dynamics results in 
non-conformal fluid dynamics (e.g. the dimensional reduction of a traceless 
stress tensor generically has non-vanishing trace). 

It is an interesting and surprising fact that the discussion of the previous
paragraph generalizes to D$p$-branes for all $p$ at the purely algebraic level.
In every case one can find a consistent truncation of the Einstein-dilaton
system which, in a purely formal manner, can be regarded as the reduction of 
negative cosmological constant Einstein gravity in a higher 
(sometimes fractional) dimension \cite{Kanitscheider:2009as}. This observation immediately yields the fluid descriptions of arbitrary D$p$-brane backgrounds as a dimensional reduction
of the conformal fluid dynamics derived in \S\ref{s:pertthy}. 

\subsection{Theories with a deconfinement transition}
\label{s:plasma}

So far, we have  studied the fluid dynamical description of field theories 
that are `deconfined' at every temperature, i.e., the free energy is ${\cal O}(N^2)$. Consider, however, a theory 
like pure Yang Mills at large $N$ which undergoes a first order deconfinement
phase transition at finite temperature. Such a system has a dual description 
in terms of a black hole only above the deconfinement temperature; the 
low temperature phase is given by a gas of glueballs, and is thermodynamically
indistinguishable from the vacuum at leading order in $N$ (free energy is ${\cal O}(1)$). 

The Scherk-Schwarz reduction of ${\cal N}=4$ Yang Mills on a circle of radius $R$ (with 
anti-periodic boundary conditions for fermions) is a simple 
example of such a theory. At strong coupling this theory undergoes a 
first order deconfinement transition at $T\, R= 2 \pi$. The gravity 
dual of the high temperature phase is simply the ${\bf S}^1$ compactification of 
the AdS$_5$ planar black hole. The gravity dual of the low temperature phase 
is a so-called AdS-soliton (a double analytic continuation of the planar Schwarzschild-AdS black hole, in which the role of time and the ${\bf S}^1$ direction are interchanged).  
At temperatures much higher than the phase transition, the effective 3-dimensional low 
energy theory is simply the dimensional reduction of the 4-dimensional conformal fluid 
system derived in earlier subsections (just as in \S\ref{s:nonconf}). 

However, at the phase transition temperature we have a new phenomenon; there
exists a new static solution of the equations of motion, a co-dimension 
one domain wall that interpolates between the AdS-soliton and the 
${\bf S}^1$ compactification of the planar AdS$_5$  black hole. Unfortunately this 
solution has been constructed only numerically \cite{Aharony:2005bm}. The domain wall is 
static in these solutions because of a pressure balance on the two sides 
(recall that the free energy, and hence the pressure, of the two phases are equal at a 
phase transition temperature). This configuration is that of a fluid 
with a boundary; the effective low energy fluctuations of this system 
consist of boundary modes (like waves on the surface of water) in addition 
to the bulk modes discussed so far in this chapter. At the ideal fluid 
level the action for boundary degrees of freedom is captured by a single 
parameter, the surface tension of the boundary (computed from the numerical solution). 

Already using the ideal 
fluid action including boundary terms, it has proved possible to construct
many stationary solutions of the fluid equations. These solutions, called 
plasma-balls and plasma-rings, have dual descriptions as black holes, black 
rings, and (in higher dimensions) black objects of more exotic topology \cite{Lahiri:2007ae}. 
The effective action for surface degrees of freedom has not been worked out
at higher orders in the derivative expansion, and appears to be an 
interesting exercise.  

One particularly interesting static solution of the equations of ideal 
fluid dynamics including boundary terms is the plasma-tube; a configuration
consisting of a domain wall that interpolates from the vacuum to the high 
temperature phase at the phase transition temperature, followed by a second  
parallel domain wall at separation $L$ that interpolates back to the vacuum. 
Such a fluid configuration is the two dimensional analogue of a 3-dimensional 
cylindrical tube of fluid, and is well known to undergo a famous fluid 
dynamical instability (to droplet formation) called the Rayleigh instability. 
For real fluids such as water, the endpoint of the Rayleigh instability is a 
series of disconnected droplets. Now the gravitational dual of the plasma tube
is an infinitely long black string in 5-dimensional gravity. This solution has the 
well known Gregory-Laflamme instability [Ch.GL] which, apparently, is dual to the 
Rayleigh instability in the long wavelength limit. The boundary 
dual of a series of disconnected droplets, on the other hand, is a series of 
disconnected black holes. This discussion at least strongly suggests that 
the end point of the Gregory-Laflamme instability consists of localized 
black holes \cite{Cardoso:2006ks}. Note that the fluid description breaks down near the `pinch off' point; the actual description of topology change in this process requires 
the use of the full field theory (e.g.\ details of interactions between water 
molecules in the case of water). 

%
\subsection{Charged fluids and anomalies}
\label{s:}

Under the AdS/CFT correspondence a global symmetry in the boundary maps to a 
gauge symmetry in the bulk. This suggests that there should 
be a duality between the long wavelength asymptotically AdS planar black hole 
solutions of the Einstein-Maxwell theory (with negative cosmological constant) and the equations of charged fluid 
dynamics. This  is a useful extension as
fluids of interest in experimental situations conserve one or more 
$U(1)$ charges in addition to energy and momentum. For instance, the flow 
of air in the atmosphere conserves air molecule number.

It is conceptually straightforward to generalize the set-up of 
section \S\ref{s:pertthy} to the study of locally thermalized charged planar black holes. 
The starting point is the construction of spacetimes  that tubewise approximate 
Reissner-Nordstr\"om AdS black hole solutions with locally varying
temperature, chemical potential and velocity. A perturbation expansion 
entirely analogous to the one outlined in \S\ref{s:pertthy} then constructs
the gravitational solutions to the Einstein-Maxwell-Chern-Simons theory 
(which forms a consistent truncation of IIB supergravity on AdS$_5 \times 
{\bf S}^5$) dual to charged fluid flows order by order in the derivative expansion \cite{Erdmenger:2008rm,Banerjee:2008th}.  This procedure also determines the equations of charged fluid dynamics to  order by order in the derivative expansion. The results of this analysis  turned out to throw up a surprise purely from the viewpoint of charged first 
order fluid dynamics, as we now explain. 

The equations of charged fluid dynamics are the conservation of the charge 
current 
\begin{equation}\label{ccc}
\nabla_{\! a} \, J^a=0
\end{equation}
together with the conservation of the stress tensor \eqref{stc}. 
Concrete fluid dynamical equations require constitutive relations that 
express the field redefinition invariant parts of the stress tensor and 
charge current in terms of expressions of first order in the derivative 
of fluid fields. The charge current for charge density $q$ takes the form
\begin{equation}\label{ccf}
J^a= q \, u^a + J^a_{diss} \ ,
\end{equation}
where $J^a_{diss}$ represents the contribution of terms with one or more 
derivatives of the fluid fields to the charge current, and is to be viewed as the 
charge current analogue of $\Pi^{ab}$; similarly to \eqref{pexp} we 
expand
\begin{equation}\label{jexp}
J^a_{diss}= \sum_{n=1}^\infty \ell_{\rm m}^n \, j_{(n)}^a \ .
\end{equation} 

Standard textbook analyses assert that 
the most general allowed form at first order for the constitutive relations 
of a relativistic charged fluid are the first equation of \eqref{constraints} along with
\begin{equation}\label{consCan1}
P^a_{\ c}\,  
\left( j^c_{(1)} + \frac{q}{\rho + P} (u_b \,\Pi_{(1)}^{bc}) \right) = \kappa \, V_1^a \ , \quad V_1^a\equiv -P^{ab} \nabla_{b}\frac{\mu}{T} + \frac{F^{ab}u_{b}}{T}
\end{equation}
while the second equation of \eqref{constraints} gets modified to
\begin{equation}\label{consCan2}
\frac{1}{d-1}\, \Pi_{(1)}^{ab} P_{ab} - \frac{\partial P}{\partial\rho} \, ( u_a \,u_b \,\Pi_{(1)}^{ab})+\frac{\partial P}{\partial q} (u_a j_{(1)}^a )  = - \beta \, \nabla_{\! c} \, u^c \ .
\end{equation}
In \eqref{consCan1} $F^{ab}$ is the non-dynamical background electromagnetic field that 
couples to the $U(1)$ current $J^a$ in \eqref{ccc} and $\mu$ is the 
chemical potential of the fluid. Provided that
$$ \eta \geq 0, ~~~ \kappa \geq 0, ~~~ \beta \geq 0 ,$$
these expressions are consistent with the positivity of divergence of 
the fluid entropy current 
\begin{equation}\label{canentc} 
J_{can}^a = s \, u^{a} - \frac{1}{T} \, u_{b} \, \Pi_{(1)}^{ab} - \frac{\mu}{T} \, j^{a}_{(1)} \,,
\end{equation}
using the alleged relation
\begin{equation}\label{diventcanc}
 \nabla_{a} J^a_{can} = -\nabla_{a} \left( \frac{u_{b}}{T} \right) \Pi_{(1)}^{ab} 
- \left( \nabla_{a} \left( \frac{\mu}{T}\right) - \frac{F_{ab}u^{b}}{T}\right) j^{a}_{(1)}.
\end{equation}

However it was found by explicit computation that the fluid dual to  
the asymptotically AdS-Einstein-Maxwell-Chern-Simons system has constitutive
relations that differ from those of \eqref{consCan1} in the following 
fashion: the r.h.s.\ of \eqref{consCan1} includes new 
terms proportional to fluid vorticity $\omega^a$ and rest frame magnetic field $B^a$ where
$$\omega^{a} = 
\frac{1}{2}\epsilon^{ab c d} \, u_b \, \partial_{c} u_{d} \ , \qquad B^{a} = \frac{1}{2}\epsilon^{ab c d} \, u_b \, F_{cd}\ . $$

In a beautiful paper \cite{Son:2009tf} pointed out the reason for the 
appearance of these new terms. When the $U(1)$ current 
has a global $U(1)^3$ triangle anomaly (as is true of the field theory dual to a 
bulk system with a 5-dimensional Chern-Simons term), \eqref{diventcanc} has an 
additional term on its r.h.s.\ proportional to this anomaly. This term spoils
the positivity of the divergence of the canonical entropy current in the 
presence of such a field. It is however consistent with the positivity 
of the divergence of a modified entropy current provided that modifications
are also made to the r.h.s.\ of  \eqref{consCan1}. 
More concretely, positivity of the entropy current in every conceivable 
circumstance requires that, in addition to the first equation of \eqref{constraints} and \eqref{consCan2}, 
\begin{equation}\label{conssSS}
\begin{split}
J^a_S&=J^a_{can}+ \sigma_{\omega} \, \omega^{a} + \sigma_B \, B^{a}\\
P^a_{\ c}\,  
\left( j^c_{(1)} + \frac{q}{\rho + P} (u_b \,\Pi_{(1)}^{bc}) \right) 
&= \kappa \, V_1^a  + \tilde{\kappa}_{\omega} \, \omega^a 
+\tilde{\kappa}_B \, B^a 
\end{split}
\end{equation}
where 
\begin{equation}\label{res}\begin{split}
\sigma_{\omega}&=c \,  \frac{\mu^3}{3T} + T  \, \mu  \, k_2 + T^2 \,  k_1\\
\sigma_B&=c  \, \frac{\mu^2}{2T} + \frac{T}{2}  \, k_2\\
\tilde{\kappa}_{\omega}&=c \left( \mu^2 -\frac{2}{3}  \, \frac{q}{\rho +P}  \, \mu^3 \right)  + T^2 \left(1-\frac{2q}{\rho +P}  \, \mu \right) k_2 - \frac{2q}{\rho +P}  \, k_1   \\
\tilde{\kappa}_{B}&=c\left( \mu -\frac{1}{2} \, \frac{q}{\rho_n+P}  \, \mu^2 \right) - \frac{T^2}{2}\frac{q}{\rho+P}  \, k_2 \\
\end{split}
\end{equation}
and $k_1$ and $k_2$ are integration constants. Further imposition of CPT invariance forces $k_2$ to vanish.

This explanation accounts for the additional transport coefficients 
in the AdS/CFT duality, but applies more generally to every fluid flow 
with a $U(1)^3$ anomaly. The effect of these new transport coefficients may
well turn out to have experimentally measurable effects in the relativistic heavy ion collisions 
 or in neutron or quark stars.

\subsection{Holographic superfluid hydrodynamics}
\label{s:}

It was pointed out by \cite{Gubser:2008px} that charged asymptotically 
AdS$_5$ planar black holes are sometimes unstable in the presence of charged scalar fields.  The endpoint of this instability is a hairy black hole: a black hole 
immersed in a charged scalar condensate. The AdS/CFT correspondence maps 
the hairy black hole to a phase in which a global $U(1)$ 
symmetry is spontaneously broken by the vacuum expectation value of a charged 
scalar operator (see [Ch.CM] for further discussion). In condensed matter physics a phase with a spontaneously  broken global $U(1)$ symmetry is referred to as a superfluid. 

The variables of relativistic 
superfluid dynamics consist of two velocity fields, the normal fluid velocity 
$u^a$ and a superfluid velocity field $u^a_s$,  together with a temperature
and chemical potential field. The superfluid velocity is the unit vector 
in the direction of $-\xi_{a}$ where $\xi_a$ is 
 the gradient of the phase of the scalar 
condensate. Conservation of the stress 
tensor and charge current together with the assertion that 
$\xi_a$ is curl free constitute the equations of superfluid 
dynamics. These equations form a closed dynamical system once 
they are supplemented with constitutive relations that express  
the stress tensor, charge current and the component of $\xi_a$ 
along the normal velocity, as functions of the fluid dynamical variables. 

It has proved possible to apply the fluid/gravity map to hairy black holes
to derive the constitutive relations for holographic superfluids, with 
interesting results. The theory of perfect superfluids was worked out by 
Landau and Tisza in the 1940s. In a beautiful recent work  \cite{Sonner:2010yx} have used the equations
 of Einstein gravity to re-derive  Landau-Tisza equations for superfluids that admit a holographic description. 
The theory of first order dissipative corrections to the equations of 
Landau-Tisza superfluidity was most completely spelled out in \cite{Putterman:1974uq}. Calculations
done within the fluid/gravity framework have led to the realization that  
the 13 parameter Clark-Putterman equations derived therein miss one parameter (under the 
assumption of parity invariance for the superfluids) and 6 more parameters
(if the superfluids are not assumed to preserve parity). A completely
satisfactory framework for superfluid hydrodynamics has been developed only 
very recently \cite{Bhattacharya:2011tr}, and the fluid/gravity map has played a major role in 
this development.

\section{Relation to other developments}
\label{s:otherdev}

Having surveyed the fluid/gravity correspondence and its various applications, we finally describe connections with other approaches.

\subsection{Implications for Israel-Stewart formalism}
\label{s:}

One  useful application of the fluid/gravity correspondence is an improvement on the `causal relativistic hydrodynamics', also known as the Israel-Muller-Stewart formalism \cite{Israel:1976tn,Israel:1979wp}. To appreciate the context, recall that a conventional theory of relativistic dissipative (i.e., irreversible) hydrodynamics, which is first order in time derivatives, is described in terms of a parabolic system of differential equations, leading to instantaneous propagation of signals.  While these apparently a-causal modes lie outside of the long wavelength regime of validity of the hydrodynamical formulation as discussed in \cite{Geroch:2001xs}, they nevertheless lead to conceptual and computational problems.
To capture the dissipative physics, \cite{Israel:1976tn} observed that second order terms are needed in the entropy current.  These render the system hyperbolic, thereby 
providing a good initial value formulation. However, the particular terms added are not all the possible ones consistent with the symmetries, so as such, the construction is somewhat ad-hoc.  Indeed, \cite{Baier:2007ix} observed in the context of conformal fluid, that the terms added do not maintain conformal invariance of the system, manifesting the incompleteness of the approach.
 The fluid/gravity construction in effect prescribes the correct completion to render the full system causal, as well as manifestly consistent with the symmetries.  We expect that due to the gravitational dual, causality will be guaranteed at all orders in the derivative expansion.

\subsection{The black hole membrane paradigm}
\label{s:membrane}

Perhaps the most salient feature of the fluid/gravity correspondence is the fact that the horizon dynamics (which in this case prescribes the dynamics of the entire spacetime) is governed by hydrodynamics.  At the face of it, such type of relation is not new; in fact for several decades 
relativists have explored the idea that spacetime, or important aspects thereof like black hole horizons, might resemble a fluid.
Early indications include black hole thermodynamics \cite{Bekenstein:1973ur,Hawking:1974sw} developed in the 70's,  analog models of  black holes \cite{Unruh:1980cg} initiated in the early 80's, and most strikingly the black hole Membrane Paradigm \cite{Thorne:1986iy,Damour:1978cg} formulated in the late-70's.  The latter realizes the idea that for external observers, black holes behave much like a fluid membrane, endowed with physical properties such as viscosity, conductivity, and so forth.  In particular, the dynamics of this membrane is governed by the familiar laws of fluid dynamics, namely the compressible Navier-Stokes equations.

Motivated by the superficial similarity between the membrane paradigm and the fluid/gravity correspondence, recently \cite{Eling:2009sj,Bredberg:2011jq} have attempted to formulate a precise derivation of the former. In \cite{Eling:2009sj} Einstein's equations in the bulk are projected onto a null hypersurface and then expanded in gradients along the hypersurface. On the other hand \cite{Bredberg:2011jq} show that one can systematically find a solution to vacuum Einstein's equations which describes the near-horizon geometry of a generic non-degenerate black hole in the long wavelength regime.

Within the fluid/gravity correspondence, the entire spacetime evolution is mapped to the dynamics of a conformal fluid, which, albeit reminiscent of the membrane paradigm, has one important twist: the membrane lives on the {\it boundary} of the spacetime (which is unambiguously defined and admits a fluid description with well-defined dynamics), and gives a perfect mirror of the full bulk physics. This  ``membrane at the end of the universe" picture is a natural consequence of the holographic nature of the fluid/gravity correspondence. 

\subsection{Blackfolds}
\label{s:blackfolds}

As in the fluid/gravity correspondence and the membrane paradigm type ideas, the blackfold approach to constructing higher-dimensional black holes 
(discussed in [ch.BF] of this book) likewise asserts that the effective theory describing the long wavelength dynamics of black hole horizons can be expressed in terms of fluid dynamics.
However, there are several important differences between these descriptions.
Since the blackfold `fluid' pertains to the effective world-volume dynamics of an extended black object as seen from far away, the intrinsic dynamics typically has to be supplemented by extrinsic dynamics, describing how the blackfold embeds in the ambient spacetime.  
 On the other hand, in the fluid/gravity correspondence the fluid resides on the boundary of asymptotically (locally) AdS spacetime, so there is no issue with extrinsic dynamics.
 
Moreover, although the blackfold formalism is most naturally formulated in asymptotically flat spacetime, by a suitable separation of scales, one can in principle consider blackfolds with any asymptotics. 
In contrast, the fluid/gravity correspondence concerns asymptotically AdS black holes.
 On the other hand, unlike all the above-mentioned approaches, fluid/gravity is the only one where there is a known physical microscopic origin to the fluid:
 it is the effective behavior of the dual field theory residing on the AdS boundary.

\section{Summary}
\label{s:summary}

The fluid/gravity correspondence provides a natural way to map solutions of fluid dynamics into those of gravity, enabling one to construct time-dependent, inhomogeneous black hole solutions to Einstein's equations, retaining full non-linearity. An interesting aspect of the construction is the manner in which classical gravity can be moulded to fit naturally with effective field theory intuition  to extract  approximate solutions. While the construction itself arose from the gauge/gravity correspondence, it is clear that it can be implemented in greater generality. 

Apart from providing interesting insights into the dynamics of gravity, the map has played an important role in clarifying various issues in fluid dynamics. The role of quantum anomalies in hydrodynamical transport, and generalizations of fluid dynamics to systems with spontaneously broken symmetries, are two examples where the fluid/gravity map has served to elucidate the underlying physics cleanly. The physical points seem much simpler to understand from the gravitational perspective; aided by this intuition one can re-evaluate the hypotheses of traditional descriptions of fluids. 

The fluid/gravity map suggests several extremely interesting technical, as 
well as conceptual, questions for the future. Some of these are:
 Does the gravitational viewpoint shed any light on 
turbulent fluid flows, or questions about singularities that develop in 
finite time from smooth initial data in fluid dynamics?
Is there a path integral formulation of fluid dynamics 
at finite $N$, and how does it map to the path integral of bulk gravity? 
Are the corrections to the classical equations of gravity constrained 
by the requirement of positivity of divergence of an `entropy current' on 
an event horizon (analogously to, and perhaps even dual to,  fluid dynamics)?
It seems likely that many interesting results remain to be discovered
in this general area.

\section{Epilogue: Einstein and Boltzmann}
\label{s:ebotlz}

As we have emphasized throughout, the equations of fluid dynamics, for which 
we have an independent field theory intuition, are dual to a long wavelength
limit of  Einstein's equations \eqref{Eeq}. It is  then natural to ask what is the field theoretical 
interpretation of the full dynamical system of equations \eqref{Eeq}? 
We believe that these equations may be conceptually  thought of as the strong 
coupling analogue of (a decoupled sector of) the Boltzmann transport equations.
 
It is well known that the linearization of the Boltzmann transport equations
about equilibrium yields an infinite set of `quasinormal modes', i.e., 
solutions to the equations of motion that all decay to zero (returning the 
system back to equilibrium) at late times. Exactly $d$ of these 
quasinormal modes are massless (in the sense that they are static in the 
infinite wavelength limit).  In textbooks on statistical 
mechanics, fluid dynamics is sometimes derived as the non-linear theory 
of this finite set of Boltzmann `quasinormal modes'. 
The remaining quasinormal modes are `fast modes' that 
decay away on a time scale set by the mean free path of kinetic theory. 

Similarly, the linearization of the equations of gravity about 
the  planar black hole has $d$ massless quasinormal modes and an infinite 
set of massive quasinormal modes. In direct analogy with the work on the 
Boltzmann equations, the fluid/gravity correspondence constructs the equations 
of fluid dynamics as the non-linear theory of these massless modes (effectively
by integrating out the massive modes, order by order). 
For this reason it is natural to think of the full set of Einstein's equations
in the presence of an event horizon
 (including all quasinormal mode degrees of freedom) 
as the strong coupling analogue of the Boltzmann transport equations. 

An important property of the Boltzmann transport equations is that they are 
irreversible; they obey the Boltzmann $H$-theorem (which asserts that a 
certain functional of kinetic variables called $H$ always increases 
in time and is maximum in equilibrium). In direct analogy, Einstein's equations, 
together with the assumption of regularity of future event horizons (and physical energy conditions),  
always obey the classic area increase theorem of general relativity. 
This suggests that the better analogy is between the Boltzmann transport 
equations and  Einstein's equations {\it plus the condition of regularity 
of the future event horizon}. The last condition breaks the time reversal 
invariance of Einstein's equations. In fact, the requirement that the future event horizon stay regular was a crucial element in our implementation of the fluid/gravity map. The 
Boltzmann theorem has a local analogue in fluid dynamics; it maps to 
the statement that the equations of fluid dynamics are accompanied 
by a local entropy current that whose divergence is everywhere non-negative. 
The area increase theorem of general relativity can be used to
construct such an entropy current for the fluid dynamics generated from the 
fluid/gravity map. 

Just like the Boltzmann equations, the system of gravitational
 equations \eqref{Eeq} can be used to study the approach to equilibrium from a highly non equilibrated starting point. In simple studies of equilibration using  
Einstein's equations \cite{Chesler:2008hg,Bhattacharyya:2009uu}, the  equilibration time, measured by the time taken for fluid dynamics to take over as the effective description, 
turns out to be extremely rapid. In more complicated 
situations the equilibration process displays sharp phase transitions 
associated with Choptuik phenomena. Indeed the equations 
\eqref{Eeq} undoubtedly contain a host of dynamical delights for the 
intrepid gravitational and statistical physicist; it seems clear that 
the fluid/gravity map is merely the tip of an iceberg's worth of connections 
between gravity and statistical physics.

\providecommand{\href}[2]{#2}\begingroup\raggedright\endgroup

\end{document}